\documentclass{aastex6}
\usepackage{gensymb}
\newcommand{\thicktilde}[1]{\mathbf{\tilde{\text{$#1$}}}}

\begin{document}


\title{X-Ray Reverberation From Black Hole Accretion Disks with Realistic Geometric Thickness}



\author{Corbin Taylor\altaffilmark{1,3} and Christopher S. Reynolds\altaffilmark{2,1,4}}


\altaffiltext{1}{Department of Astronomy, University of Maryland, 1113 Physical Sciences Complex (Building 415), College Park, MD 20742-2421, USA}
\altaffiltext{2}{Institute of Astronomy, Madingley Road, Cambridge, CB3 0HA}
\altaffiltext{3}{cjtaylor@astro.umd.edu}
\altaffiltext{4}{csr12@ast.cam.ac.uk}

\begin{abstract}

X-ray reverberation in AGN, believed to be the result of the reprocessing of corona photons by the underlying accretion disk, has allowed us to probe the properties of the inner-most regions of the accretion flow and the central black hole. This process is modeled via raytracing in the Kerr metric, with the disk thickness almost ubiquitously assumed to be negligible (razor-thin) and the corona commonly approximated as a point source located along the polar axis (a lamppost). In this work, we use the new raytracing suite, {\tt Fenrir}, to explore the effect that accretion disk geometry has on reverberation signatures, assuming a lamppost configuration but allowing for a finite disk scale height. We characterize the signatures of finite disk thickness in the reverberation transfer-function and calculate how they might manifest in observed lag-frequency spectra. We also show that a disk-hugging corona (approximated by off-axis point-like flares) exhibits characteristics that are qualitatively different from observation, thus providing further evidence for a flaring corona that is separated from the underlying disk material.

\end{abstract}




\section{Introduction} \label{sec:intro}

The study of X-ray variability in active galactic nuclei (AGN) probes the structure of and the physical processes that occur in the inner-most regions of the accretion flow. Reverberation is one such phenomenon seen in Seyfert galaxies, where the bands associated with the reflection spectrum will lag behind those dominated by the high-energy power law. The power-law is believed to be produced when a corona containing hot electrons ($\sim\,100$ keV) upscatters the thermal UV photons from the accretion disk ($\sim\,10$ keV) into the X-ray regime. While many of these photons will escape the system and produce the observed continuum, others will be reprocessed by the disk, creating the reflection spectrum and resulting in a natural path difference between direct and reprocessed photons \citep{Fabian+1989, Uttley+2014}. The associated lag allows one to probe the properties of the corona and the underlying disk, with most of the X-rays coming from a very compact region $< \, 10 \, r_{\rm g}$ from the central black hole \citep{Fabian+2015}.

Fabian et al. (1989) first proposed reflection as a possible explanation to the broad emission line observed in the stellar mass black hole binary (BHB) Cyg X-1 by \cite{Barr+1985}, arguing that the feature was consistent with that of fluorescent Fe K$\alpha$ due to reprocessing of continuum radiation, the line broadened and skewed due to Doppler and relativistic effects \citep{Cunningham1975}. \cite{Fabian+1989} noted that the reflection process would naturally produce a lag between the continuum and the reprocessed radiation, and that the wings of profile would respond prior to the centroid. The first confirmed detection of reverberation lag was in the AGN 1H 0707-495, with the soft excess at 0.3 -1.0 keV lagging behind the 1-4 keV band by $\sim \, 30$ s at $> 6\times10^{-4}$ Hz, consistent with a compact continuum source within a few gravitational radii of the event horizon \citep{Fabian+2009}. Similar lags were later observed in the broad Fe K$\alpha$ \citep{Zoghbi+2012} and the Compton hump \citep{Zoghbi+2014}, and have now been shown to be fairly common phenomena in Seyfert galaxies (\citealt{Kara+2016} and references therein).

While reverberation is the leading hypothesis for high frequency X-ray variability in AGN, at lower Fourier frequencies (e.g $\nu$ $<$ $2\times10^{-4}$ Hz in ARK 564, \citealt{Kara+2016}) the relationship changes, with continuum dominated bands lagging behind those associated with the reflection spectrum, the lag magnitude increasing with photon energy. This low-frequency hard lag was first discovered in BHB Cyg X-1 \citep{Page1985, Miyamoto+Kitamoto1989} and then later in the AGN NGC 7469 \citep{Papadakis+2001}, with linear rms-flux relationships \citep{Uttley+McHardy2001} and lognormal flux distributions \citep{Gaskell2004, Uttley+2005} seen in both system classes. The leading hypothesis to explain this phenomenon is via accretion fluctuations that propagate inwards through a disk corona, the harder lags coming from the inner-most regions of the flow \citep{Kotov+2001, Arevalo+Uttley2006}. The rms-flux relation and flux distribution have been reproduced via recent MHD simulations, being shown to be related to the dynamo. \citep{Hogg+Reynolds2016}.

Reverberation is modeled through relativistic raytracing, where one solves the equations of motion for photons in the curved spacetime around the central compact object, tracing their paths from the X-ray emitting corona to the disk, and then from the disk to the observer. The first such calculations were performed by \cite{Stella1990} and \cite{Matt+Perola1992}, and a more rigorous exploration in a Kerr geometry was performed later by \cite{Reynolds+1999}. Raytracing produces both timing and spectral information, allowing one to predict the reverberation lag behavior as a function of Fourier frequency and photon energy \citep{Nowak+1999, Uttley+2014}. There have been a few common simplifying assumptions made in these calculations, with the disk almost ubiquitously assumed to have negligible vertical structure (i.e. "razor thin") and the corona most often being approximated as a point source situated along the rotation axis (a lamppost) \citep{Martocchia+Matt1996, Reynolds+Begelman1997, Miniutti+Fabian2004}. The lamppost corona model is commonly visualized as being a hot electron population near the base of a jet \citep{Biretta+2002, Ghisellini+2004} or in the black hole magnetosphere \citep{Hirotani+Okamoto1998}, and while consistent with estimations of coronal compactness in Seyfert galaxies \citep{Reis+Miller2013, Fabian+2015}, it must be emphasized to be a fiducial model chosen due to a lack of true understanding of coronal geometry. 

While these calculations have been able to reproduce much of the high-frequency variability and the time-averaged spectral characteristics observed in BHB and AGN \citep{Cackett+2014, Emm+2014, Chainakun+Young2015}, these simplified models still do not account for all aspects of the observed behavior, such as the low-frequency hard lag and the dip in the lag-energy spectrum at $\sim 3$ keV observed in many active galaxies (e.g. 1H 0707-495, \citealt{Wilkins+2016}). Recent work by \cite{Wilkins+2016} and \cite{Chainakun+Young2017} expands upon the lamppost model, allowing for extended coronal geometries and for coherent fluctuations through the corona, thus unifying the modeling of high-frequency and low-frequency variability; these models have shown promise, being successfully fit to XMM observations of I Zw 1 \citep{Wilkins+2017}. While there had been early exploration of the effects accretion disk geometry may have on the time-averaged reflection spectrum (e.g. disk self-eclipsing, \citealt{Pariev+Bromley1998, Wu+Wang2007}), these unified reverberation models still rely on the assumption of a razor-thin disk.

Internal pressures within the accretion disk would naturally result in non-zero scale heights, which may not be negligible compared to the other physical scales relevant to the problem (e.g. the height of the lamppost), especially in super-Eddington flows where the radiative efficiency is believed to be small in some models (see \citealt{Jiang+2017} and references therein). In \cite{Taylor+Reynolds2018}, we introduced a new raytracing suite ({\tt Fenrir}) that allows for more complex accretion disk geometries, thus further expanding upon the simple reflection model. Assuming the disk to have finite thickness consistent with a classic optically thick, geometrically thin, radiation pressure dominated \cite{Shakura+Sunyaev1973} accretion disk, and using the lamppost as a fiducial model, we explored the effects that disk thickness may have on the predicted time-averaged spectrum. Focusing on mass accretion rates roughly consistent with moderately bright Seyfert galaxies [$\dot{M}/\dot{M}_{\rm Edd}\,\in\,\{0.1, 0.2, 0.3\}$], we compared the spectra from {\tt Fenrir} to that which is predicted from the razor-thin approximation ({\tt RELXILL}, \citealt{Garcia+2014, Dauser+2014}). We found that, at the razor-thin limit, {\tt Fenrir} is consistent with other contemporary models, but the spectral models start to diverge significantly when $\dot{M} \, > \, 0.1 \, \dot{M}_{\rm Edd}$. Predominantly, these changes could be attributed to "self-shielding", where the inner edges of the disk (that are gravitationally redshifted) act to shield the outer regions, resulting in a suppression of the blue peak and an overall shifting of the line towards lower energies. With these results, we concluded that accretion disk geometry should not be neglected in the detailed modeling of moderate-to-high luminosity AGN reflection spectra, and thus it is reasonable to explore consequences of finite disk thickness in reverberation.

In this work, we follow-up the results of \cite{Taylor+Reynolds2018} by exploring the effects that a finite thickness has on reverberation signatures, using {\tt Fenrir} to calculate lag as a function of Fourier frequency and photon energy. We present the transfer functions, lag-frequency spectra, and lag-energy spectra using the lamppost approximation and the previous disk model, as well as the case where the corona has been positioned off-axis, rotating with the disk and situated at some small height above the disk surface (a rough approximation to a magnetic reconnection event close to the surface). We explore the lag signatures associated with this "disk-hugging" corona, and ask if it would be possible for such a corona to mimic a lamppost once disk thickness is taken into account.

In Section 2, we give a brief summary of the {\tt Fenrir} raytracing suite, make explicit our simplifying assumptions, and explain the cross spectrum formalism upon which our analysis is based. In Section 3, we present the results for the lamppost approximation, followed by an exploration of the "disk-hugging" corona. In Section 4, we discuss possible consequences that disk thickness may have on the estimation of model parameters, and in Section 5 we give a brief summary of our results and possible future work.

\section{Methods} \label{sec:methods}

{\tt Fenrir} \citep{Taylor+Reynolds2018} calculates the reverberation properties of AGN and BHB by integrating null-geodesics through Kerr spacetime \citep{Kerr1963, Bardeen+1972} that is described by Boyer-Lindquist coordinates \citep{Boyer+Lindquist1967}, tracing their trajectories from an X-ray emitting corona to the disk and from the disk to an observer positioned at a radial distance of $r\,=\,10^{3}\,r_{\rm g}$ at some angle $\theta \, = \, i$ measured from the black hole rotational axis. For the rest of this work, unless specified otherwise, we use units of $r_{\rm g}$ $\equiv$ $GM/c^{2}$ (gravitational radius) and $c$ (the speed of light). In standard cgs units, $r_{\rm g}$ $\sim$ $1.5M_{6}\times10^{11}$ cm $\sim$ $M_{6}\times10^{-2}$ AU and $r_{\rm g}$/$c$ $\sim$ $5M_{6}$ s, where $M_{6}$ is the mass of the compact object in units of $10^{6}\,M_{\odot}$.

\begin{figure*}
\centering
\includegraphics[width=0.48\linewidth]{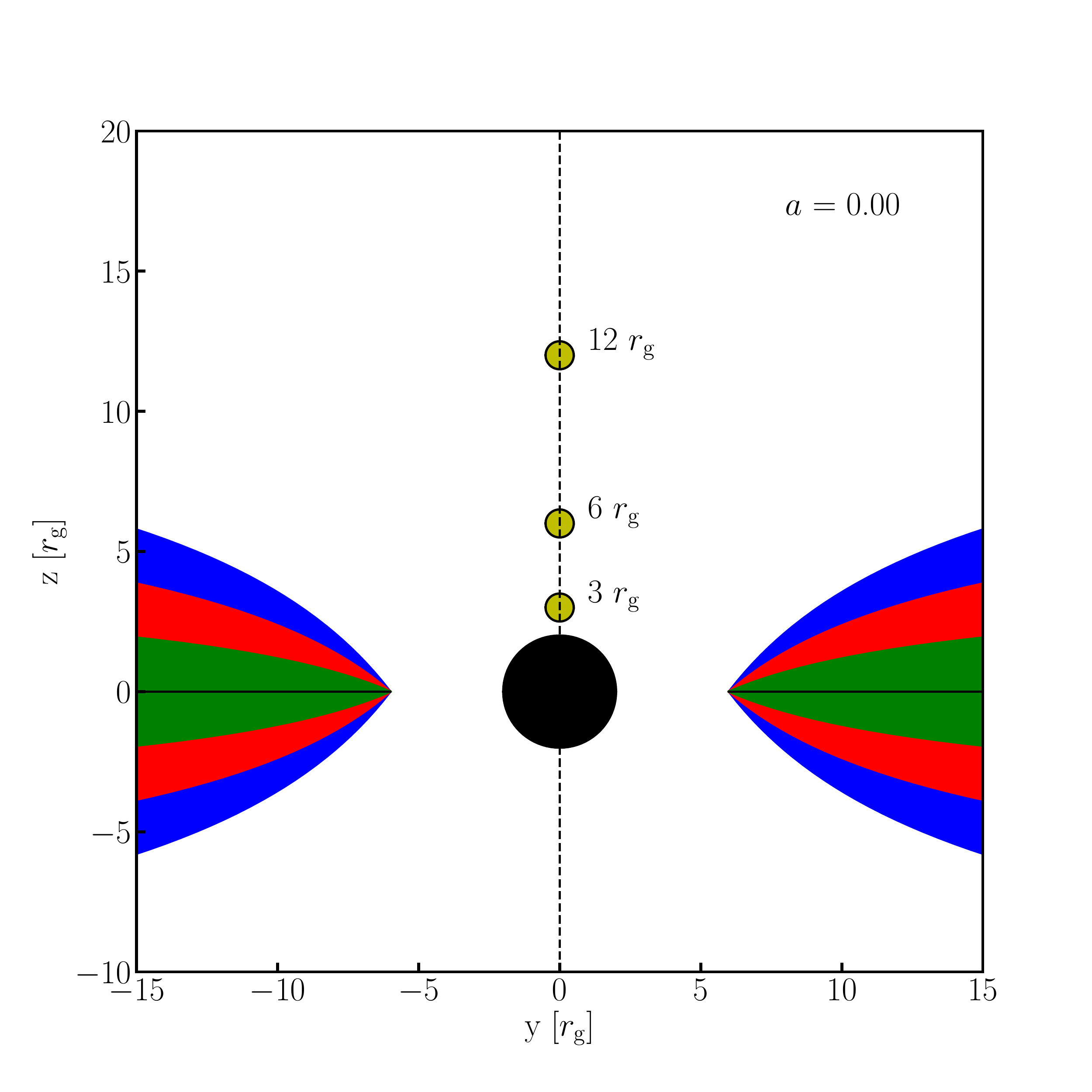}
\includegraphics[width=0.48\linewidth]{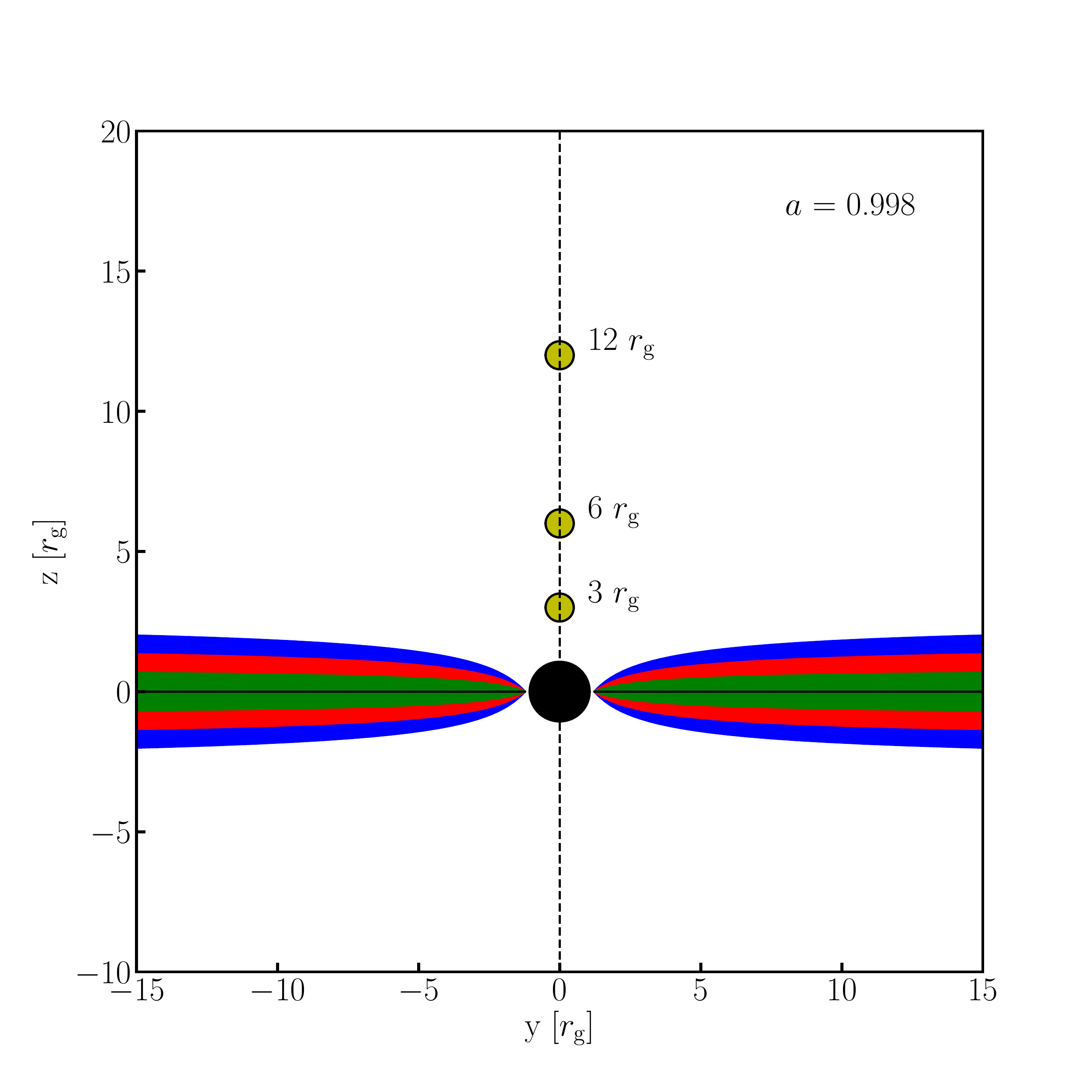}
\caption{Illustrations (expanded from Figure 1 of \citealt{Taylor+Reynolds2018}) of non-spinning ($a$ = 0.00, left) and rapidly-spinning ($a$ = 0.998, right) black holes, each with examples of a razor-thin disk (black) and finite-thickness disks with a half-thickness given by Equation \ref{eq:ss73} and a mass accretion rate of $\dot{M}$ = 0.1 (green), 0.2 (red), and 0.3 (blue). Also included are three example lamppost coronae with corresponding heights of $h$ = 3, 6, and 12 $r_{\rm g}$. In general, for a given spin, increasing $\dot{M}$ increases the thickness of the disk, while for a given accretion rate, increasing $a$ will decrease disk thickness. As one can see, for coronae that are close to the event horizon, the disk thickness can be of a similar scale to that of the $h$, and thus may not be negligible when modeling X-ray reverberation in AGN.}
\label{cartoons}
\end{figure*}

Accretion disk thickness is incorporated by using the disk surface as a stopping condition. In this work, for a black hole of spin $a \, \equiv \, Jc/GM^{2}$ (where $J$ is the angular momentum), we have assumed the disk to be an optically thick, geometrically thin, and radiation-pressure dominated \citep{Shakura+Sunyaev1973}, the reflecting surface defined by a half-thickness $z(\rho)$ equal to twice the pressure scale height,

\begin{equation}
\centering
z(\rho) = \frac{3}{\eta}\left(\frac{\dot{M}}{\dot{M}_{\rm Edd}}\right)\left[1 - \left(\frac{r_{\rm ISCO}}{\rho}\right)^{\frac{1}{2}}\right] \, \, r_{\rm g}
\label{eq:ss73}
\end{equation}
where $\rho \, \equiv \, r\sin\theta$ is defined to be the cylindrical radius, $r_{\rm ISCO}$ is the radius of the innermost stable circular orbit, and $\dot{M}/\dot{M}_{\rm Edd}$ is the Eddington accretion ratio. Allowing $E(r_{\rm ISCO})$ to denote the total specific energy of a massive particle in a circular orbit in the mid-plane at $r_{\rm ISCO}$, then $\eta$ = 1 - $E$($r_{\rm ISCO}$) is the radiative efficiency \citep{Bardeen+1972}. The disk is assumed to rotating as a series of co-centric solid cylinders: an element situated at the disk surface at some $\rho$ orbits the central compact object with the same 3-velocity as the matter in the midplane ($\theta = \pi/2$), which is assumed to be in prograde Keplerian orbits with coordinate angular velocity $\Omega_{\rm K}$ = $(\rho^{\frac{3}{2}} + a)^{-1}$ \citep{Cunningham1975}. The components of the disk element four-velocity are then constructed from this coordinate angular velocity ($U^{t}$/$U^{\phi}$ = $\Omega_{\rm K}$) and Lorentz invariance (i.e. $U^{\nu}U_{\nu}$ = -1). Figure \ref{cartoons} (expanded from Figure 1 in \citealt{Taylor+Reynolds2018}) presents the cases of a Schwarzschild ($a$ = 0.00, left) and rapidly-spinning ($a$ = 0.998, right) black hole, each with various disk models: a razor-thin disk (black) and finite-thickness disk corresponding to the mass accretion rates of $\dot{M}$ = 0.1 (green), 0.2 (red), and 0.3 (blue) $\dot{M}_{\rm Edd}$. Each panel illustrate three lamppost coronae (see below) at heights of $h$ = 3, 6, and 12 $r_{\rm g}$ along the black hole's polar axis. While more complex geometries are able to be implemented within the {\tt Fenrir} framework, we have chosen this simple form as an illustrative case, a more complete exploration being beyond the scope of this work.

Generalizing the methodology outlined in \cite{Wilkins+Fabian2012}, the corona is approximated by an isotropic point source that flashes instantaneously, orbiting the black hole with velocity four-vector $\textbf{U}$ = ($U^{t}$, 0, 0, $U^{\phi}$), the components of which are determined by the specific coronal model chosen. We calculate the conserved photon quantities for the first integration by constructing an orthonormal tetrad \{$\textbf{e}_{(t)}$, $\textbf{e}_{(r)}$, $\textbf{e}_{(\theta)}$, $\textbf{e}_{(\phi)}$\} of a frame that is instantaneously at rest with the corona,

\begin{equation}
\centering
\eta_{(\alpha)(\beta)} = g_{\mu\nu}e^{\nu}_{(\alpha)}e^{\mu}_{(\beta)}, \, \textbf{e}_{(t)} = \textbf{U}
\label{eq:tetrad}
\end{equation}
where $\eta_{(\alpha)(\beta)}$ is the Minkowski metric tensor.  From this, the photon energy ($E$), $\phi$ angular momentum ($l$), and Carter constant ($Q$) in Boyer-Lindquist coordinates can be found (see the appendix of \citealt{Wilkins+Fabian2012}).  If $p_{(\alpha)}$ is the photon momentum one-form in the instantaneous rest frame and $E$ is the photon energy, then $p_{\mu} = p_{(\alpha)}e^{(\alpha)}_{\mu}$ is the momentum one-form in the Boyer-Lindquist coordinates and

\begin{equation}
\centering
l = p_{\phi}, \, \, \, \, Q = p_{\theta}^{2} - a^{2}E^{2}\cos^{2}\theta + l^{2}\cot^{2}\theta.
\label{eq:momentum}
\end{equation}
For the case of a corona in a locally non-rotating frame (LNRF, Bardeen et al. 1972), these equations reduce to those quoted in \cite{Reynolds+1999} and \cite{Karas+1992}. For a lamppost, one situates the corona in a LNRF along the polar axis ($\theta$ = 0) and some height $r$ = $h$. The off-axis case is orbiting at the velocity of the disk ($U^{\phi}$/$U^{t}$ = $d\phi$/$dt$ = $\Omega_{\rm K}$) at a cylindrical radius ($\rho_{\rm c}$), located at a height $h_{\rm d}$ above the surface of the disk, and instantaneously at some azimuthal angle $\phi_{\rm c}$.

Defining the moment at which the direct coronal flare is seen by the observer as $t\, =\, 0$, one can describe the observed light curve of a single flash event with its subsequent reflection as,

\begin{equation}
\centering
F(E,t) = F_{\rm C}(E,t) + F_{\rm R}(E,t)
\label{eq:lightcurve1}
\end{equation}
where $F_{\rm C}$ and $F_{\rm R}$ is the observed photon flux from the corona and reflection respectively. $F_{\rm R} (E,t)$ is described by a 2D transfer function $\Psi (E,t)$, which gives the observed response from the disk as a function of energy and time \citep{Reynolds+1999} and is calculated from the output of {\tt Fenrir}. If one approximates the corona flash as a $\delta$-function in time with an associated continuum power-law $\propto \, E^{-\Gamma}$ where $\Gamma$ is the photon number index, then the normalized coronal component $F_{\rm C} (E,t)$ can be written as,

\begin{equation}
\centering
F_{\rm C}(E,t) = \frac{1}{R}\left(\frac{F_{\rm R,tot}}{F_{\rm C,tot}}\right)E^{-\Gamma}\delta(t), \, \, \, F_{\rm C,tot} = \int^{\infty}_{t = 0}\int^{10 keV}_{E = 0.1 keV} E^{-\Gamma}dEdt, \, \, \, F_{\rm R,tot} = \int^{\infty}_{t = 0}\int^{10 keV}_{E = 0.1 keV} \Psi(E,t)dEdt 
\label{eq:lightcurve2}
\end{equation}
where $F_{\rm R,tot}$ and $F_{\rm C,tot}$ are the un-normalized time-and-energy integrated reflection and coronal photon fluxes, and $R$ is the average reflection fraction. We have chosen an energy range equal to that of XMM-Newton (0.1-10 keV) for calculating $F_{\rm R,tot}$ and $F_{\rm C,tot}$. It must be noted that, while in truth AGN variability is much more complex than a single $\delta$-function flash followed by delayed response from the disk, $F(E,t)$ can be described a sum of such events if there is a linear correspondence between the reflected and coronal fluxes, as would be the case from a pure reflection scenario absent of further complexities.

Taking the Fourier transform of $F(E,t)$, $\thicktilde{F}(E,\nu)$, one can construct the cross spectrum [$C(E,\nu)$] with respect to some corona-dominated reference energy $E_{0}$,

\begin{equation}
\centering
C(E,\nu) = \thicktilde{F}^{*}(E_{0},\nu)\thicktilde{F}(E,\nu), \, \, \Delta t(E,\nu) = \frac{\arg[C(E,\nu)]}{2\pi\nu}
\label{eq:cross-spec}
\end{equation}
where $\thicktilde{F}^{*}$ is the complex conjugate of $\thicktilde{F}$ and $\Delta t$ is the time lag between $E$ and $E_{0}$ as a function of Fourier frequency $\nu$. From Equations \ref{eq:lightcurve1} and \ref{eq:lightcurve2}, one finds,

\begin{equation}
\centering
\Delta t (E,\nu) = arg\left[1 + RE^{\Gamma}\left(\frac{F_{\rm C,tot}}{F_{\rm R,tot}}\right)\thicktilde{\Psi}(E,\nu)\right]/2\pi\nu.
\label{eq:lag}
\end{equation}
Note here that $arg[C(E,\nu)]$ is only dependent on the ratio of the magnitude of the separate components rather than their absolute magnitudes. For this work, we have assumed an reflection fraction of $R$ = 1, which was chosen as a fiducial value that is consistent within an order of magnitude with much of the literature (e.g. 1H 0707-495, \citealt{Wilkins+2016}). While it is true that $R$ is physically meaningful and much recent work has been dedicated to it (e.g. \citealt{Dauser+2014}), it is not the primary focus of this study.

From $\Delta t(E,\nu)$, one derives the lag-frequency spectrum [$\Delta t(\nu)$] by integrating $\Delta t(E,t)$ over a large energy band of interest, while the lag-energy spectrum [$\Delta t(E)$] is calculated by taking the average lag across a specified frequency range for a series of energy bins \citep{Uttley+2014}. While these quantities are commonly used in analyzing reverberation signatures, it must be emphasized that the lag-frequency and lag-energy spectra are only cross sections of the complete set of information carried by a given signal, and thus treating them in isolation is inherently limiting. A full cross-spectrum analysis would be ideal, but such work is still in its infancy \citep{Chainakun+Young2015, Bachetti+Hupp2018, Mastroserio+2018} and is beyond the scope of this work.

\begin{figure*}
\centering
\includegraphics[width=0.48\linewidth]{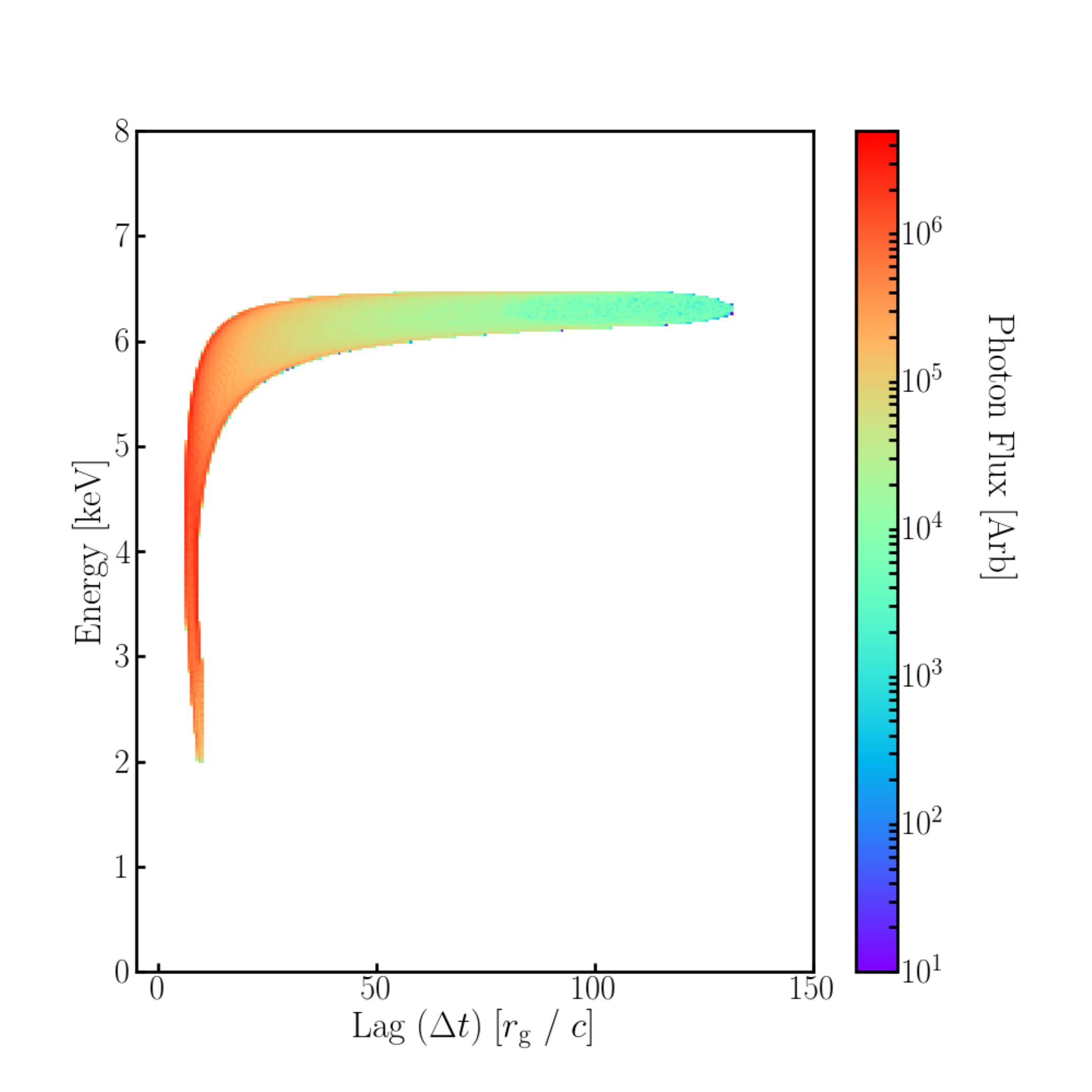}
\includegraphics[width=0.48\linewidth]{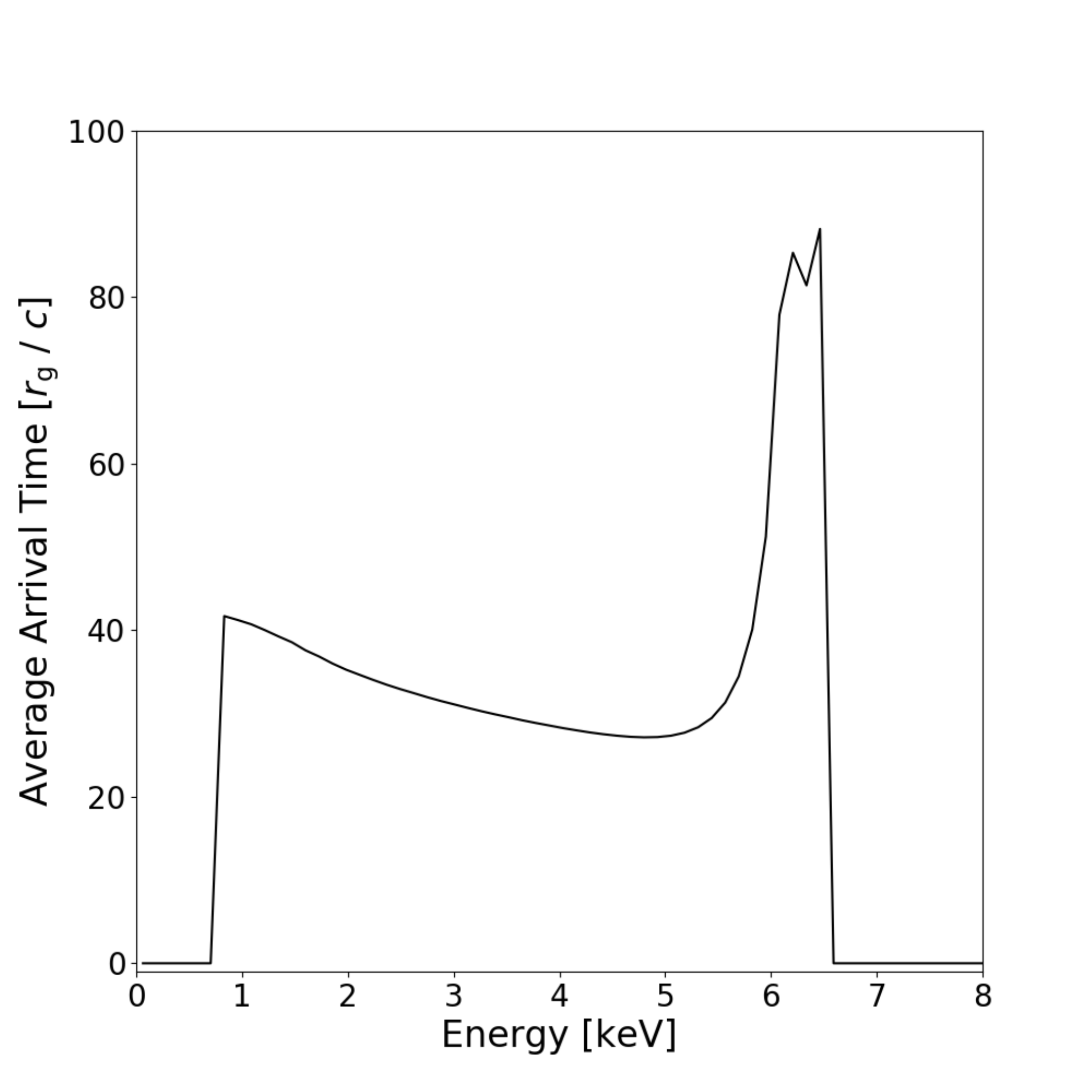}
\includegraphics[width=0.48\linewidth]{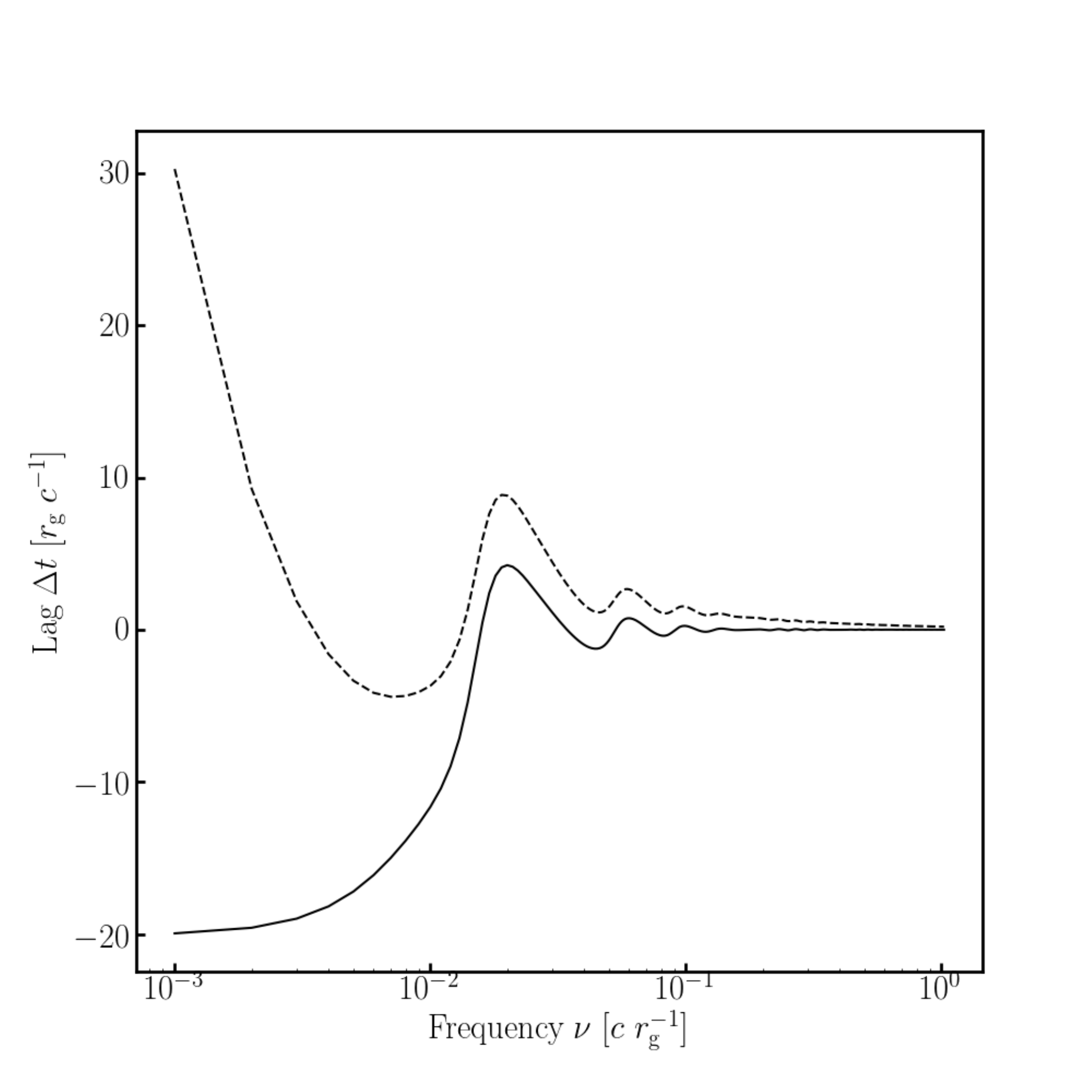}
\includegraphics[width=0.48\linewidth]{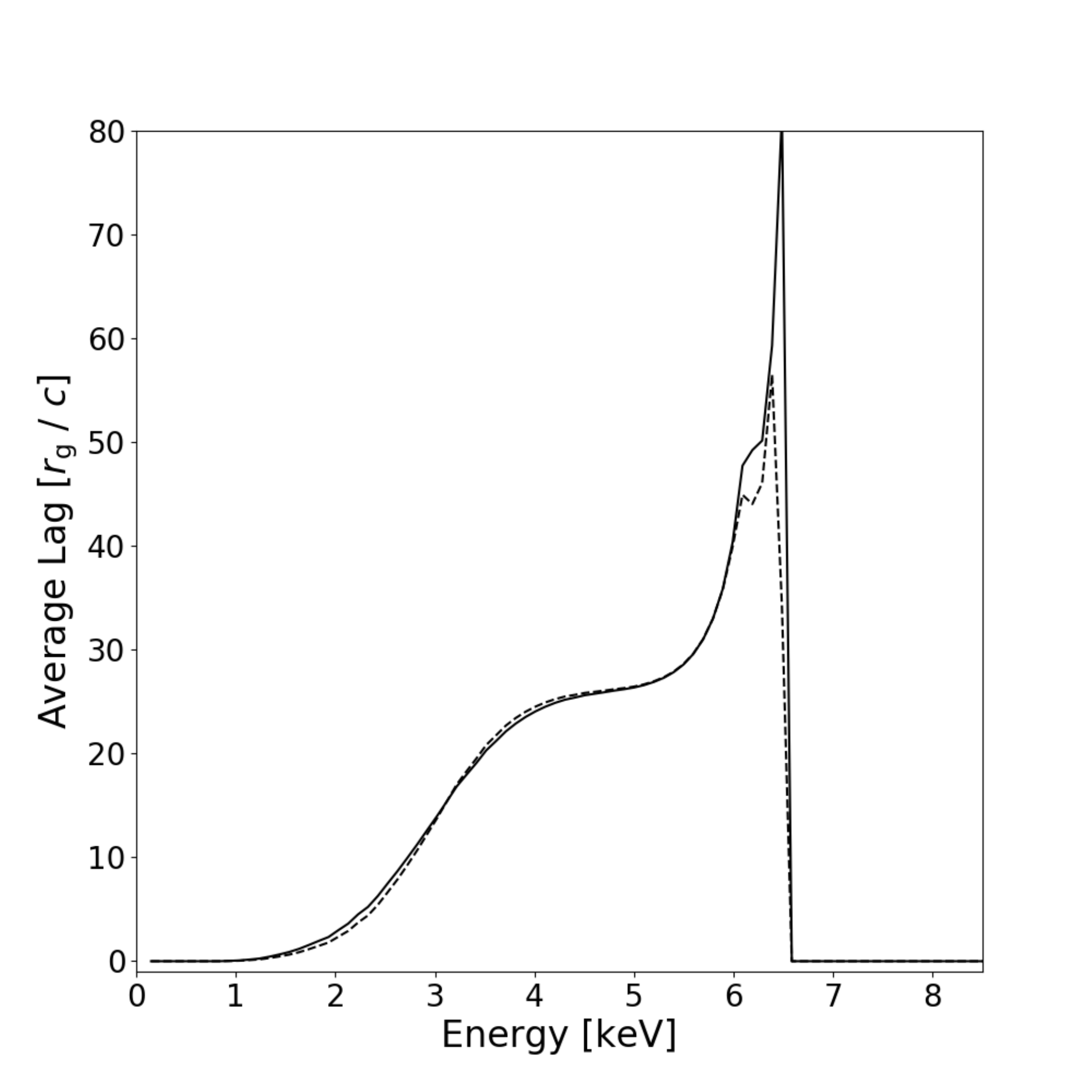}
\caption{(Top left) Example transfer function $\Psi(E,\Delta t)$ for the case of a rapidly spinning black hole ($a$ = 0.99) with a lamppost corona at $h$ = 12 $r_{\rm g}$ and a razor-thin accretion disk, observed at an angle $i$ = 15\degree away from the polar axis, along with the associated average arrival time as a function of energy (top right) absent the Fourier formalism and dilution. We have assumed that the disk is neutral, with a Fe K$\alpha$ fluorescence line with a rest energy of 6.4 keV. (Bottom left) The lag-frequency spectra and lag-energy spectra (bottom right) associated with the case above. In the left panel, we have included the lag-frequency spectrum that includes the low-frequency hard lag (dashed) and one where it is omitted (solid), where we have modeled the hard lag as a power-law. The lag-energy spectra have been created by averaging the time-dependent lag over the frequency ranges $\Delta \nu$ = $1-3\times10^{-3}\,c r_{\rm g}^{-1}$  (solid) and $5-8\times10^{-3}\,c r_{\rm g}^{-1}$ (dashed), where the differences are due to the outer disk responding at lower frequencies due to the travel time.}
\label{examples1}
\end{figure*}

Figure \ref{examples1} (upper left) gives an example transfer function [$\Psi(E,\Delta t)$] produced from {\tt Fenrir} for the case of a rapidly spinning black hole ($a = 0.99$) with a neutral razor-thin accretion disk [$E_{\rm rest}(Fe K\alpha)$ = 6.4 keV] that is being irradiated by a lamppost corona at $h = 12\, r_{\rm g}$. Approximately $\sim \, 24 r_{\rm g}/c$ after the direct coronal X-rays react the observer (designated to be $t = 0$), the observer receives the initial reflected light followed by photons from two divergent wings, one corresponding to some spread around the rest frame energy (a "blue wing") and one corresponding to increasingly lower energies (a "red wing"). The phenomenology of these features is explored in \cite{Reynolds+1999}: the initial lag and the blue wing can be thought of as due to the natural difference in the light travel time between the direct coronal radiation and the reflected light, with larger lags representative of photons being reprocessed by material at ever greater radii.The red wing however is a consequence of relativistic nature of the problem, with strong gravity resulting in radiation being reprocessed from the inner regions of the disk to be redshifted and time delayed (Shapiro delay, \citealt{Shapiro1964}). This, along with the extra time it takes for photons to reach the inner disk as they spiral around a spinning black hole from frame dragging, results in the emitting region appearing to be an inward-moving, ever-reddening annulus in the rest frame of the observer. One can see this relationship between photon arrival time and energy in the upper-right panel, where one sees that, consistent with the transfer function, the greatest lags being at low (the red wing) and high (the blue wing) energies.

Figure \ref{examples1} also the presents lag-frequency spectrum for the same system (lower left), where we have chosen to use the common convention that a negative lag represents the case of the energy of interest ($E$) lagging behind the reference energy $E_{0}$, or vice versa.  The solid line represents the reflection-only spectrum while the dashed line is the complete spectrum, with the low-frequency hard lag modeled as a power-law. As one can see, at frequencies of $\sim 8\times10^{-3}$ $c$/$r_{\rm g}$, $E$ lags $E_{0}$, consistent with the reverberation paradigm (high-frequency soft lags), while at lower frequencies, $E_{0}$ lags $E$ (low-frequency hard lags). The positive spike at approximately $2\times10^{-2}$ $c$/$r_{\rm g}$ followed by the oscillation of the spectrum about $\Delta t$ = 0 is due to phase wrapping \citep{Uttley+2014}. The corresponding lag-energy spectra is presenting in the lower right panel, having been averaged over the frequency ranges $\Delta \nu$ = $1-3\times10^{-3}\,c r_{\rm g}^{-1}$  (black) and $5-8\times10^{-3}\,c r_{\rm g}^{-1}$. (grey dashed). The lag-energy spectrum changes with frequency, with the high energy peak of the profile being suppressed at high frequencies. This is a direct result of lag being the result of light travel time: the observed photons that are close to the rest energy come from annuli at large radii, and thus further away from the irradiating corona than the material emitting photons at greater redshifts. Note that the difference between the upper and lower right panels are naturally explained by dilution, where the flux from the corona at $\Delta t$ = 0 has resulted in the decrease in the lag-signal, the magnitude of this suppression naturally being greater at lower energies due to the slope of the coronal continuum.

\section{Results}\label{sec:results}

Using the methods described in Section \ref{sec:methods}, we calculated the transfer functions, lag-frequency, and lag-energy spectra using two separate coronal models: \a lamppost and a "disk-hugging" corona (approximated as both a single off-axis flash and as an annulus). The disk thickness, given by Equation \ref{eq:ss73}, has been chosen to correspond to a mass accretion rate $\dot{M} \in \{0.1\,\dot{M}_{\rm Edd}, 0.2\,\dot{M}_{\rm Edd}, 0.3\,\dot{M}_{\rm Edd}\}$. We have also performed these same calculations for a razor-thin disk for comparison.

\subsection{The On-Axis Lamppost Corona}\label{sec:lamppost}

For the lamppost corona, each case has a corresponding coronal height ($h$) along the polar axis, along with a black hole spin ($a$), an inclination angle ($i$) relative to the polar axis, and a mass accretion rate ($\dot{M}$). We chose the values for these parameters explored previously in \cite{Taylor+Reynolds2018}, where $a \in \{0.00, 0.90, 0.99\}$, $h \in \{3\,r_{\rm g}, 6\,r_{\rm g}, 12\,r_{\rm g}\}$,  $i \in \{15\degree, 30\degree, 60\degree \}$, and  $\dot{M} \in \{0.1\,\dot{M}_{\rm Edd}, 0.2\,\dot{M}_{\rm Edd}, 0.3\,\dot{M}_{\rm Edd}\}$. As explained in our previous work, these values were chosen to be a broad (albeit coarse) sampling of parameter space that would be of interest in the study of Seyfert 1 galaxies \citep{Reynolds2014}.

\begin{figure*}
\centering
\includegraphics[width=0.48\linewidth]{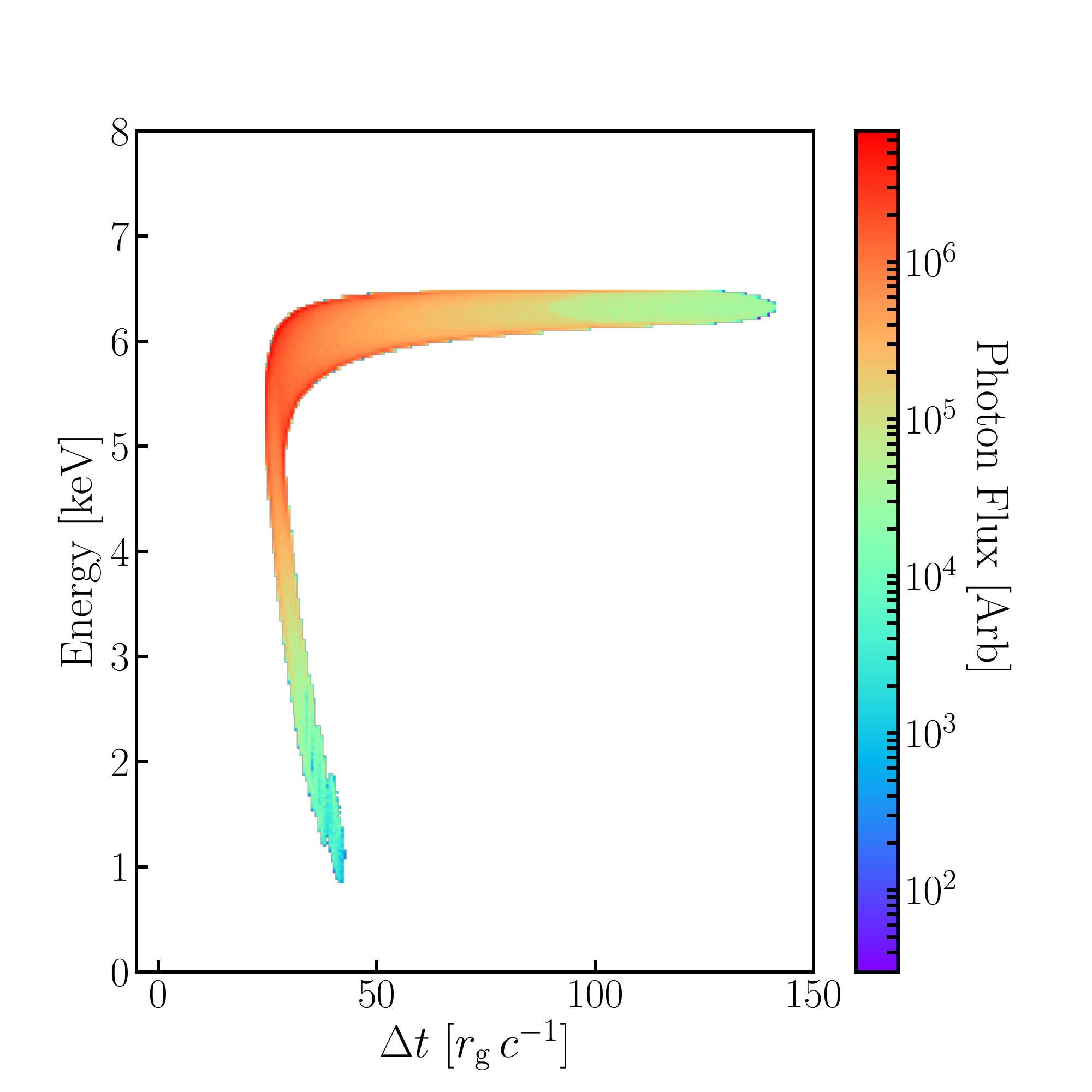}
\includegraphics[width=0.48\linewidth]{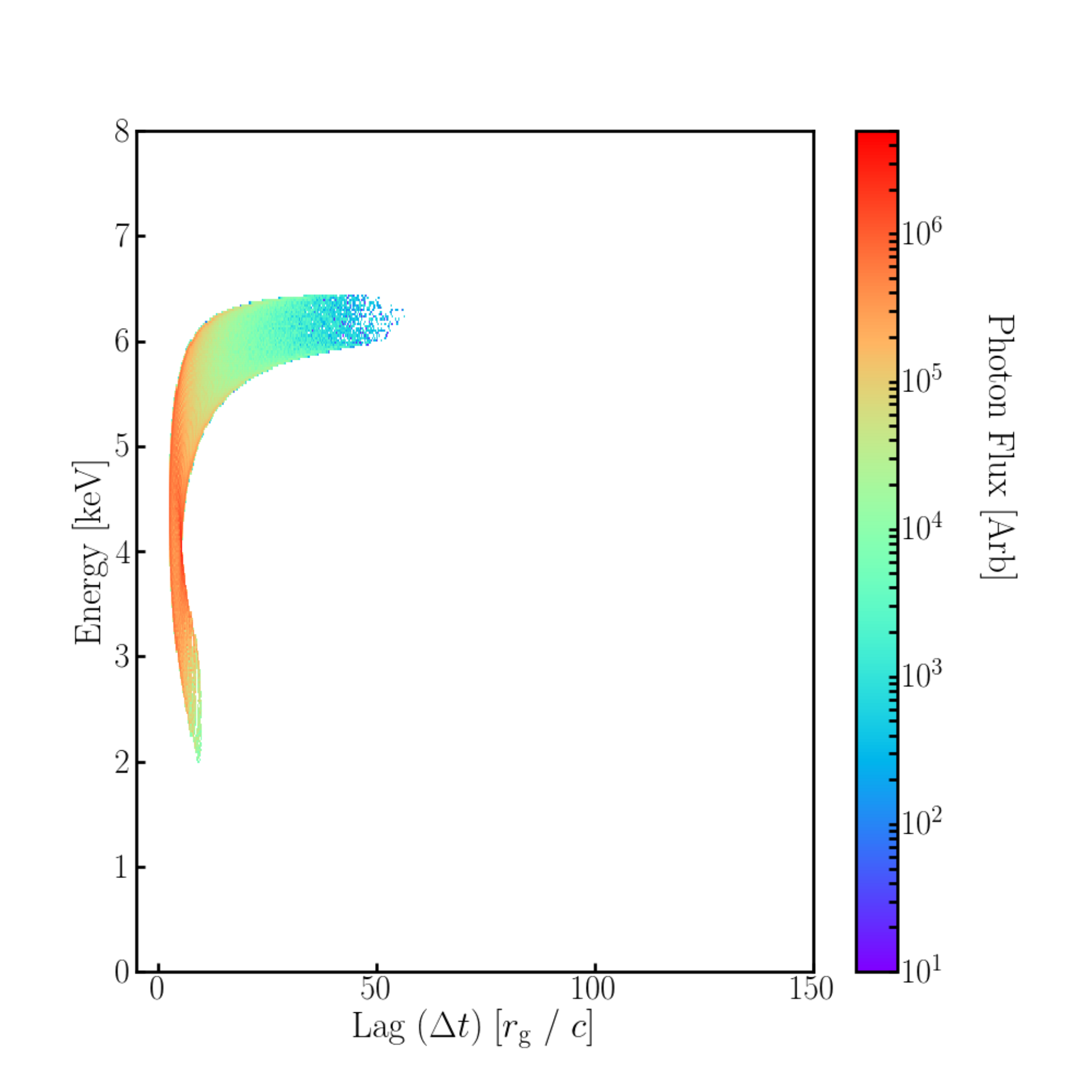}
\includegraphics[width=0.48\linewidth]{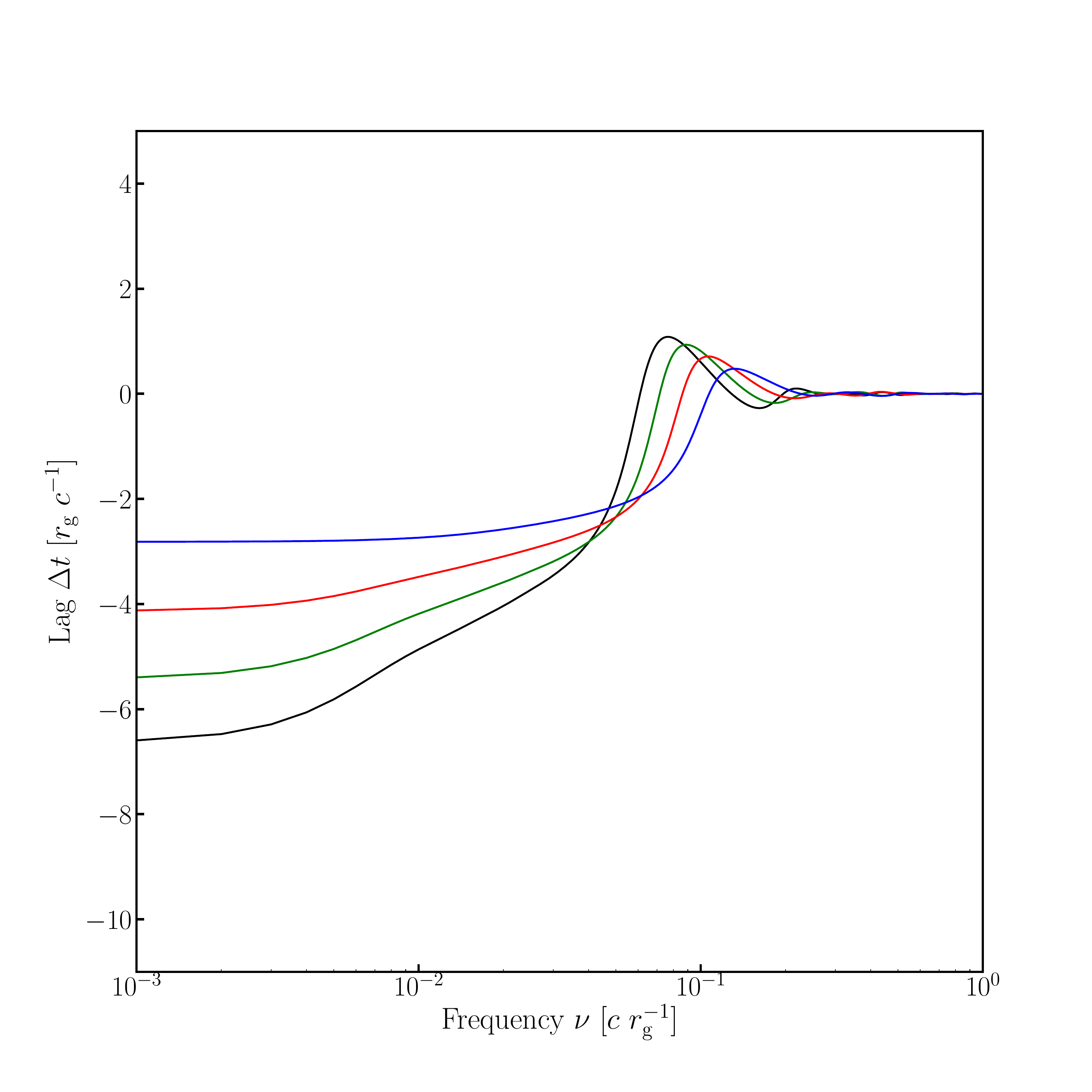}
\includegraphics[width=0.48\linewidth]{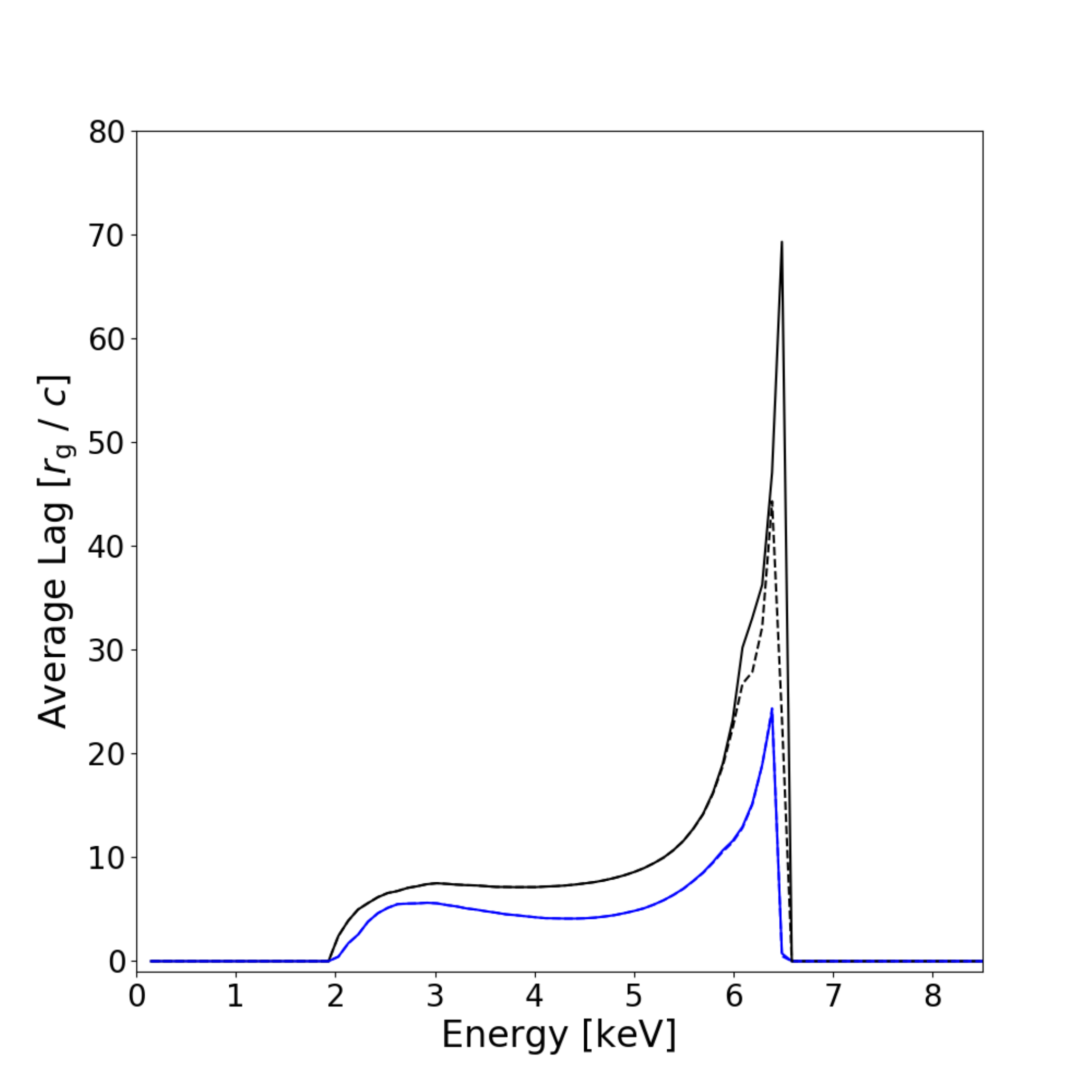}
\caption{(Top) The transfer functions [$\Psi (E, \Delta t)$] for the case of a moderately-spinning black hole ($a$ = 0.90), observed at an angle $i$ = 15\degree, with a lamppost corona situated at a height of $h$ = 3 $r_{\rm g}$. We have used two different disk geometries: a razor-thin accretion disk (left) and a finite thickness disk with a half-thickness given by Equation \ref{eq:ss73} and a mass accretion rate $\dot{M}$ = 0.3 $\dot{M}_{\rm Edd}$ (right). When one allows the disk thickness to be appreciable compared to $h$, the late-time "blue wing" ($\Delta t$ $>$ 50 $r_{\rm g}$/$c$) is truncated due to self-shielding: the inner edges of the disk act to shield the outer material from being irradiated by the corona. (Bottom) The lag-frequency (left) and lag-energy (right) spectra corresponding to these parameter values. In the lag-frequency spectra, we present the razor-thin disk (black), as well as the finite thickness case with $\dot{M}$ = 0.1 (green), 0.2 (red), and 0.3 (blue) $\dot{M}_{\rm Edd}$, which clearly show that the overall lag decreases with increasing $\dot{M}$, while the phase wrapping frequency increases and magnitude decreases with increasing $\dot{M}$. In the lag-energy spectra, we show the razor-thin disk (black) and finite thickness disk with $\dot{M}$ = 0.3 $\dot{M}_{\rm Edd}$ (blue) for the frequency ranges of $\nu$ = $1-3\times10^{-3} \, c/r_{\rm g}$ (solid line) and $5-8\times10^{-3} \, c/r_{\rm g}$ (dashed line). As one can see, the increase in disk thickness results in an overall decrease in the lag, while also causing the lag-energy spectra in both frequency bands to become indistinguishable. This is due to the suppression of the late-time blue wing resulting in less lag at higher energies, while a decreased travel time from corona to disk has decreased the travel time at lower energies. Overall, this results in a natural "sharpening" of the response in the time domain, thus resulting in a broad signal in frequency space and the observed similarity of the lag-energy relation across different frequency bands.}
\label{transfers1}
\end{figure*}

\begin{figure*}
\centering
\includegraphics[width=0.48\linewidth]{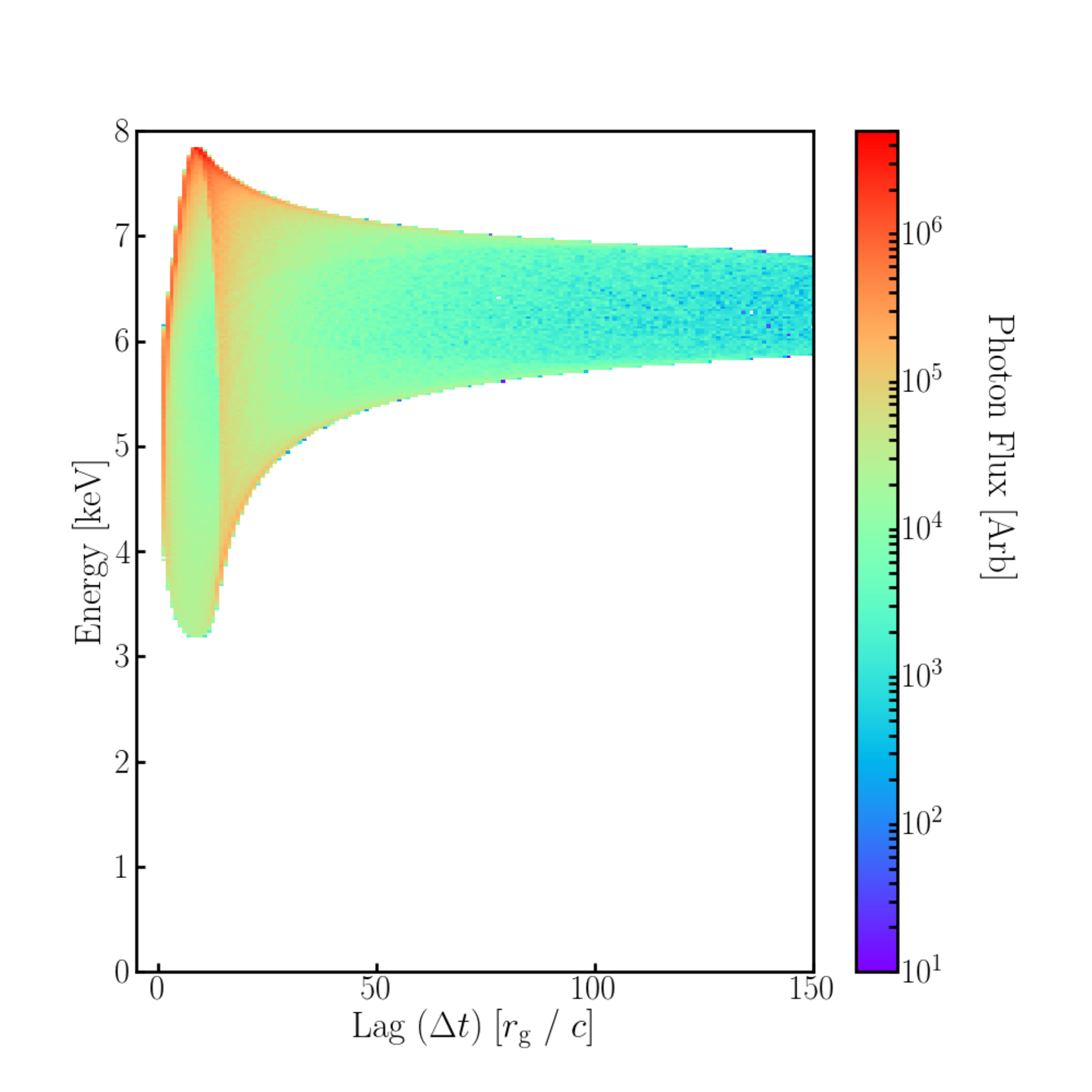}
\includegraphics[width=0.48\linewidth]{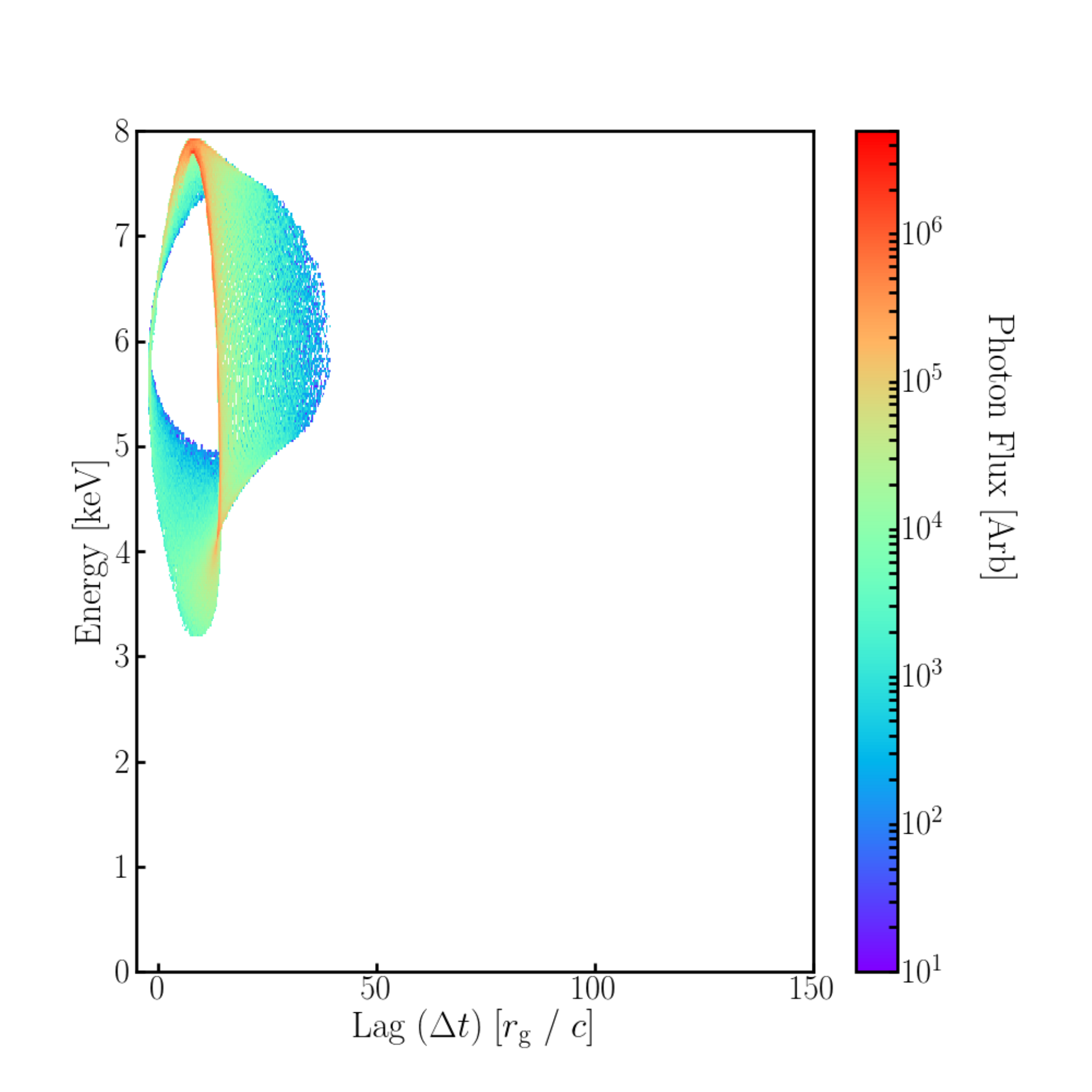}
\includegraphics[width=0.48\linewidth]{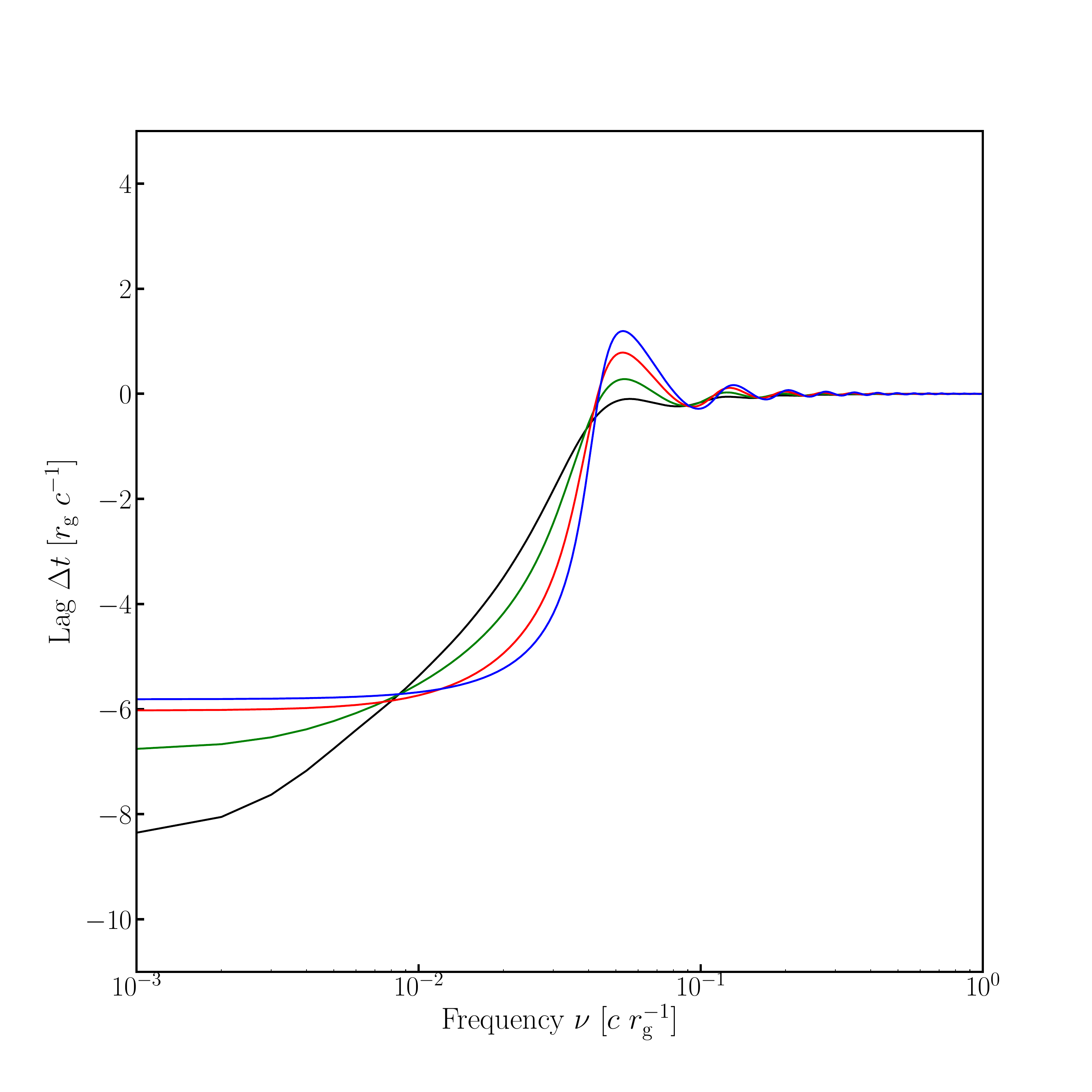}
\includegraphics[width=0.48\linewidth]{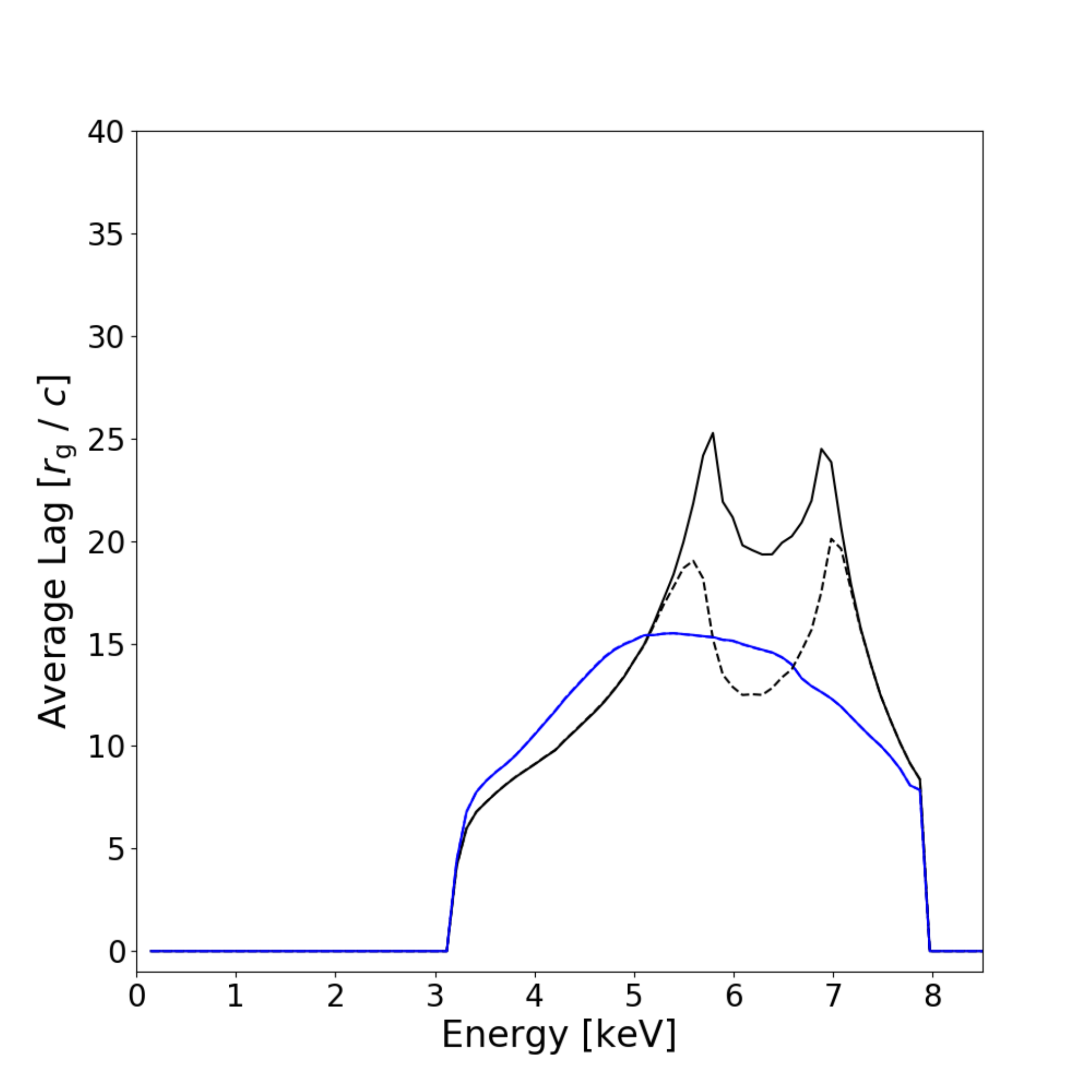}
\caption{Same as Figure \ref{transfers1}, but with a Schwarzschild black hole ($a$ = 0.00), observed at an angle $i$ = 60\degree, with a lamppost corona at $h$ = 3 $r_{\rm g}$. As before, we see that a disk with finite thickness (right) has a truncated late-time "blue-wing" compared to the transfer function calculated using a razor-thin accretion disk (left). We also see a "hollowing" of the broad feature at $\Delta t$ $<$ 25 $r_{\rm g}$/$c$, where self-shielding has prevented part of the front side of the disk ($\rho$ $\sim$ 50 $r_{\rm g}$) from being irradiated by the corona. The lag in the lag-frequency spectra is likewise decreased with increasing $\dot{M}$, but we see now that there is a very dramatic increase in the slope, and an overall increase in the phase wrapping magnitude. As in the previous figure, we see a similarity in the finite-thickness case between frequency bands, but now the centroid of the lag-energy profile has shifted towards lower energies.} 
\label{transfers2}
\end{figure*}

\begin{figure*}
\centering
\includegraphics[width=0.48\linewidth]{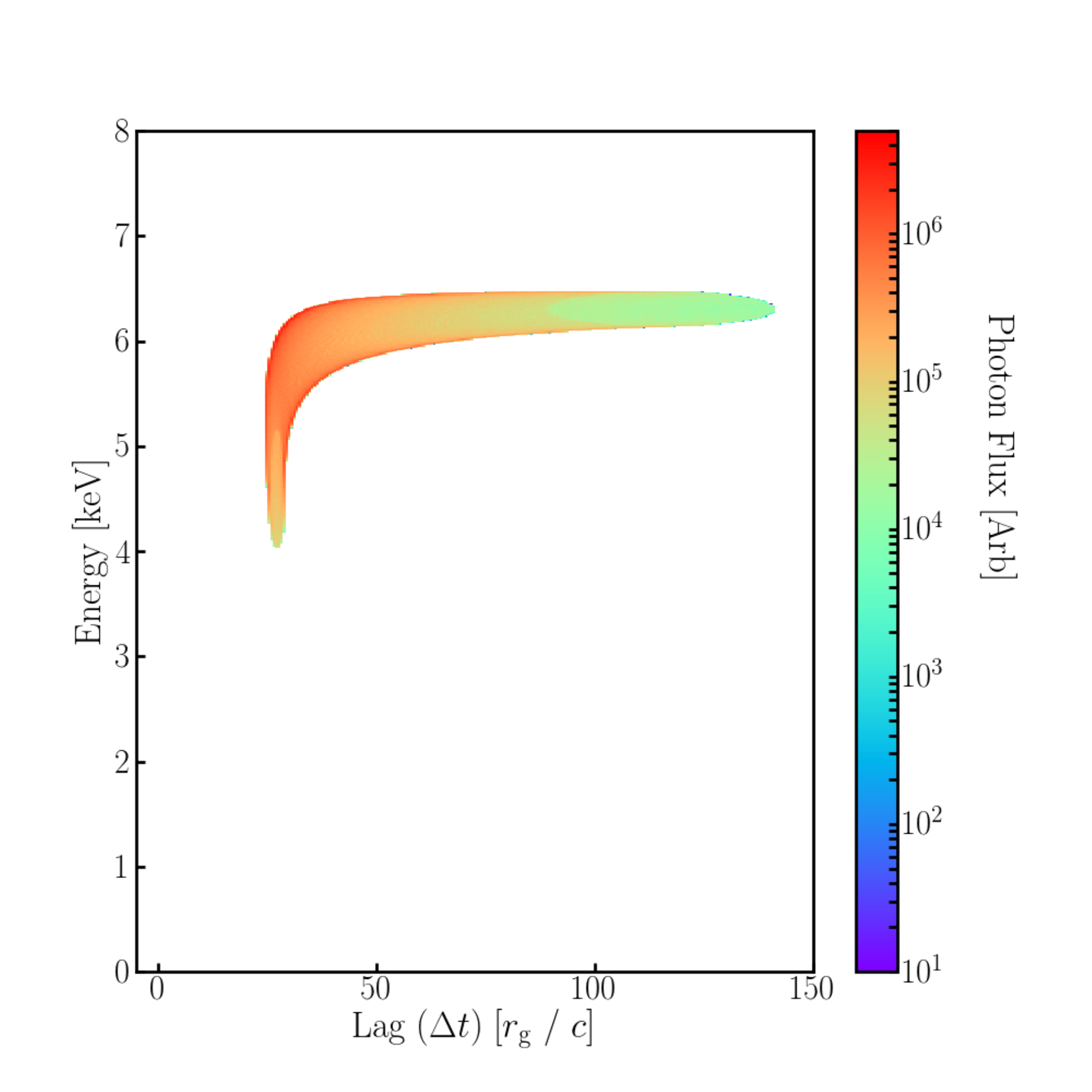}
\includegraphics[width=0.48\linewidth]{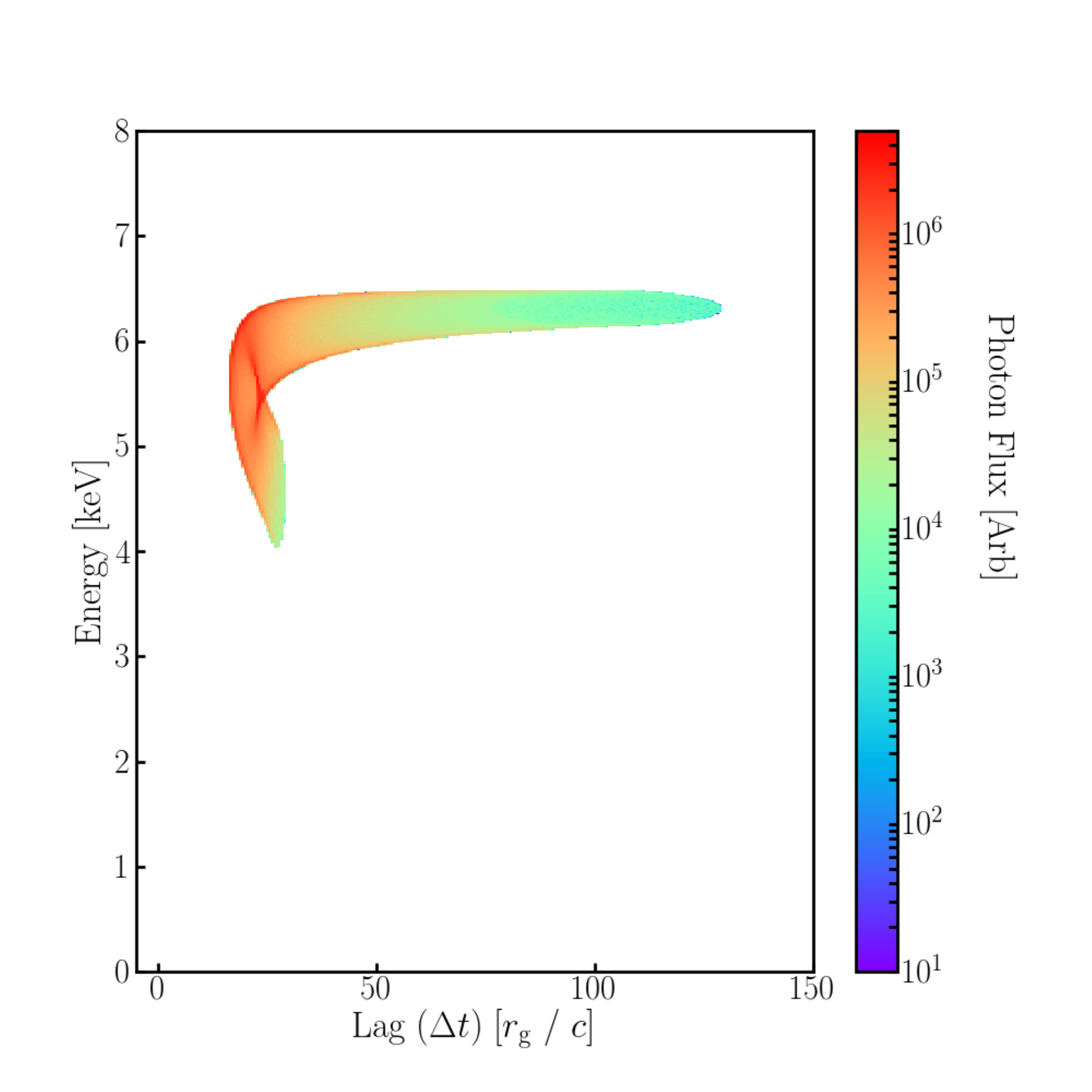}
\includegraphics[width=0.48\linewidth]{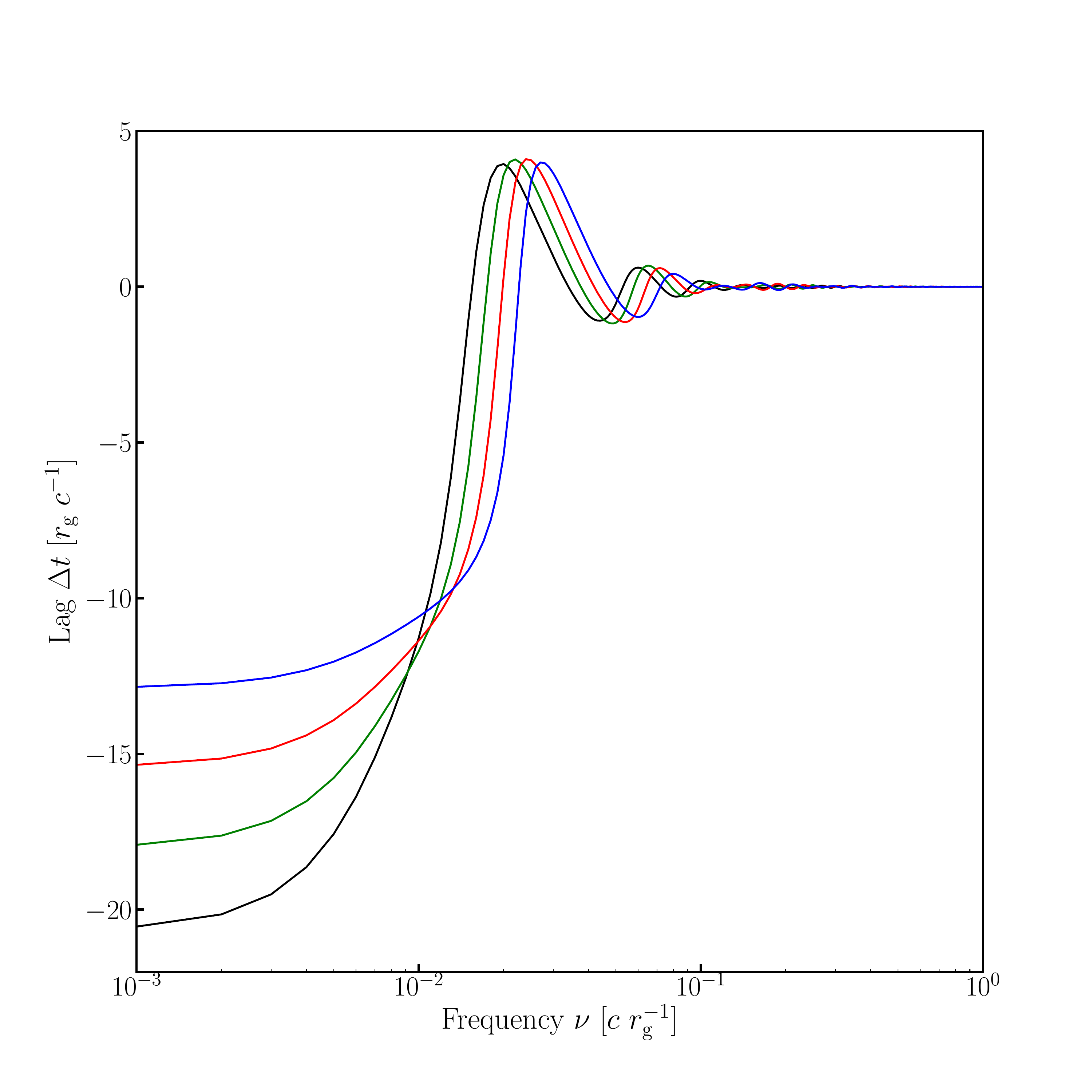}
\includegraphics[width=0.48\linewidth]{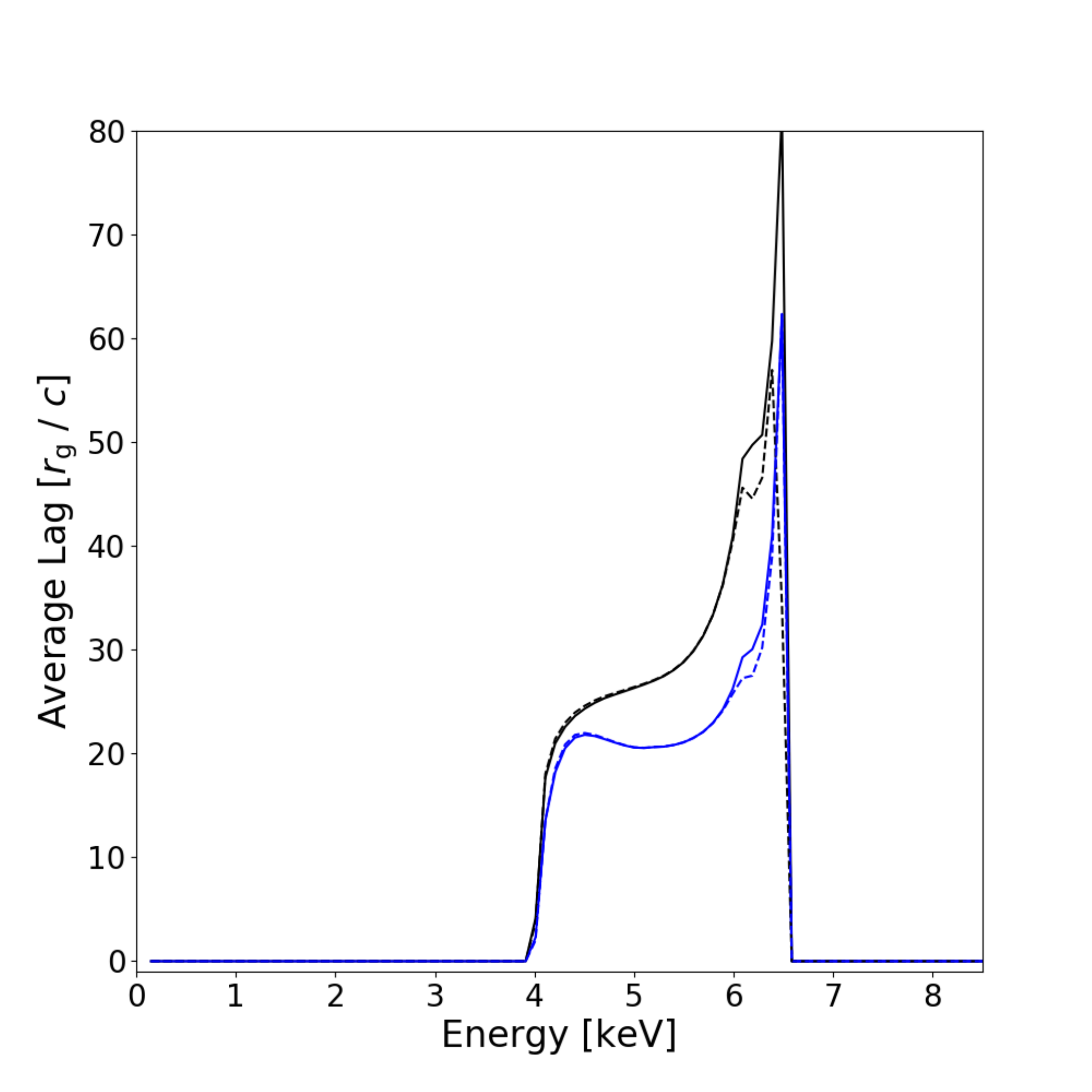}
\caption{Same as Figure \ref{transfers1}, but with a Schwarzschild black hold ($a$ = 0.00), observed at an angle $i$ = 15\degree, with a lamppost corona at $h$ = 12 $r_{\rm g}$. The larger distance of the corona above the disk results in a lack of self-shielding, however the geometry of the inner disk has resulted in the "red-wing" changing shape. In particular, finite scale height shortens the path length of photons from the corona to the disk, shortening the initial lag, and as the minimum energy is dictated by $r_{\rm ISCO}$ (which is the same in both models), this results in the apparent change of slope. In the lag-frequency spectra and lag-energy spectra, we see a decrease in the overall lag, and an increase in the phase wrapping frequency, but not a corresponding increase or decrease in phase wrapping magnitude. The fact that $h$ is large has made disk self-shielding much less relevant, and thus one sees a variation with frequency in lag-energy spectra corresponding to the finite-thickness disk.} 
\label{transfers3}
\end{figure*}

Figures \ref{transfers1}, \ref{transfers2}, and \ref{transfers3} give three pairs of transfer functions as examples of the three most prominent effects disk thickness has on these functions, each pair presenting the control case of a razor-thin accretion disk (left) and the case of a finite-thickness disk (right) with $\dot{M}$ = 0.3 $\dot{M}_{\rm Edd}$. Figure \ref{transfers1} illustrates how a non-zero vertical scale height can truncate the transfer function when $h$ is small, the late-time "blue wing" being entirely absent. This is naturally explained by disk self-shielding, where the convex inner edge of the disk blocks the outer radii from being irradiated by the X-ray corona, thus preventing substantial reprocessing at these radii and resulting in a very steep negative radial gradient; for a visual example of this, see Figures 3 \& 5 of \cite{Taylor+Reynolds2018}.

Another effect observed in our exploration is presented in Figure \ref{transfers2}, where we present the transfer functions from a Schwarzschild black hole with a corona close to the event horizon ($h$ = 3 $r_{\rm g}$), observed at a steep angle ($i$ = 60\degree). Like the previous case, disk thickness has resulted in truncation, however there is also a notable "hollowing" of the transfer function shortly after the initial response ($\Delta t$ $<$ 25 $r_{\rm g}$) at energies 5 keV $<$ $E$ $<$ 7.4 keV, centered at $\sim$ 6.4 keV (the rest energy). This effect, observed at high inclinations, is once again due to shadowing: this energy and lag time corresponds to moderate cylindrical radii ($\rho$ $\sim$ 50 $r_{\rm g}$) that reside on the side of the disk nearest the observer which, while irradiated in the razor-thin limit, are blocked from the corona due to intervening disk material.

A final prominent effect observed in the transfer functions of the lamppost cases is presented in Figure \ref{transfers3}, where there is an apparent change in the slope of the "red tail" between the razor-thin and finite-thickness accretion disk cases. In this particular case, one sees that the lag-energy slope becomes more shallow with the increasing scale height. This is due to the decrease in the initial reflection lag while the bottom of the tail remains fixed, and can easily be explained by the convex shape of the inner disk which tapers to a null scale height at $\rho$ = $r_{\rm ISCO}$. The increased scale height decreases the path difference between the photons that are observed directly from the coronal flash and the first observed reprocessed photons, thus pushing the left-most portion of the transfer function towards smaller lags. The flux from the minimum energy of the red tail, however, comes from the inner edge where the lag is dominated by time dilation and the negligible scale height would not have appreciable affect on said lag. The red tail slope is then not only a consequence of the position of the inner-most radius of the disk, but also of the geometry of the inner accretion flow.

\begin{figure*}
\centering
\includegraphics[width=0.96\linewidth]{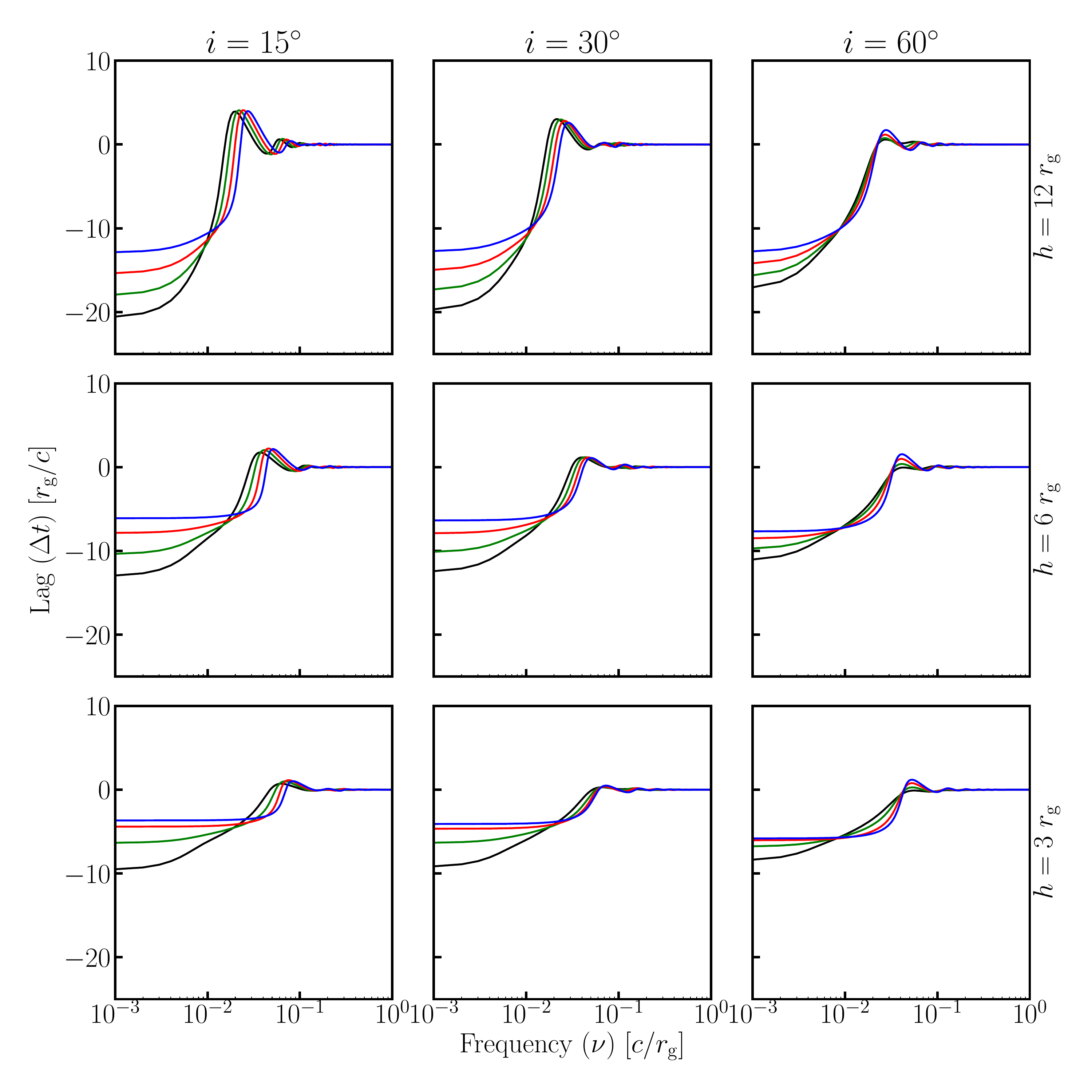}
\caption{Lag-frequency spectra for the case of a Schwarzschild black hole ($a$ = 0.00) with a lamppost corona and an accretion disk that is either razor-thin (black) or has finite-thickness $\dot{M}$ = 0.1 (green), 0.2 (red), and 0.3 (blue) $\dot{M}_{\rm Edd}$. Each row represents a different coronal height $h$ = 12 (top), 6 (center), and 3 (bottom) $r_{\rm g}$, while each column represents a different observer angle $i$ = 15\degree (left), 30\degree (center), and 60\degree (right). For each value panel, the magnitude of the negative lag at low frequencies for finite thickness disks is less than the corresponding lag in the case of a razor-thin disk, with the lag magnitude being negatively correlated with $\dot{M}$. At higher frequencies in the cases of $h$ = 3 and 6 $r_{\rm g}$, one sees that disk thickness slope of the lag-frequency spectra to steepen with increasing disk thickness. Finally, phase wrapping appears to be affected by disk thickness, with the frequency of the first peak increasing with increasing disk thickness when $i$ = 15\degree and 30\degree; there is some small variation in the magnitude of said peak. Overall, one can reasonably expect to underestimate $h$ if one were to model a system with non-negligible disk thickness (e.g. as expected in super-Eddington AGN) using the razor-thin disk approximation.} 
\label{lag-freq1}
\end{figure*}

\begin{figure*}
\centering
\includegraphics[width=0.96\linewidth]{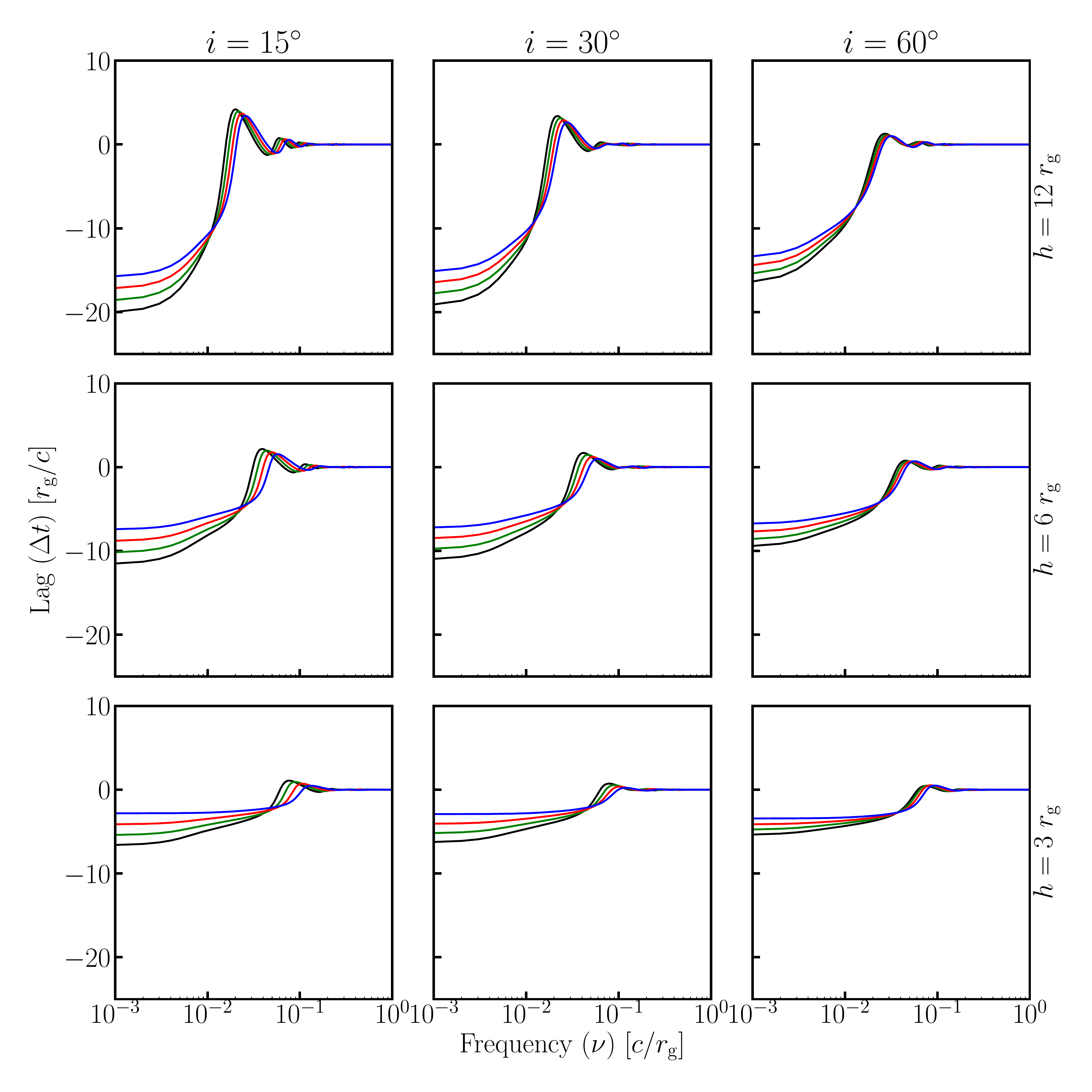}
\caption{Same as Figure \ref{lag-freq1}, but with a moderately-spinning black hole ($a$ = 0.90). Like before, the magnitude of the low-frequency negative lag and the phase wrapping frequency are both positively correlated with disk thickness. As the efficiency $\eta$ (see Equation \ref{eq:ss73}) increases with increasing $a$, the change in the negative lag magnitude is not as great due to the inverse relationship between $\eta$ and disk thickness.} 
\label{lag-freq2}
\end{figure*}

\begin{figure*}
\centering
\includegraphics[width=0.96\linewidth]{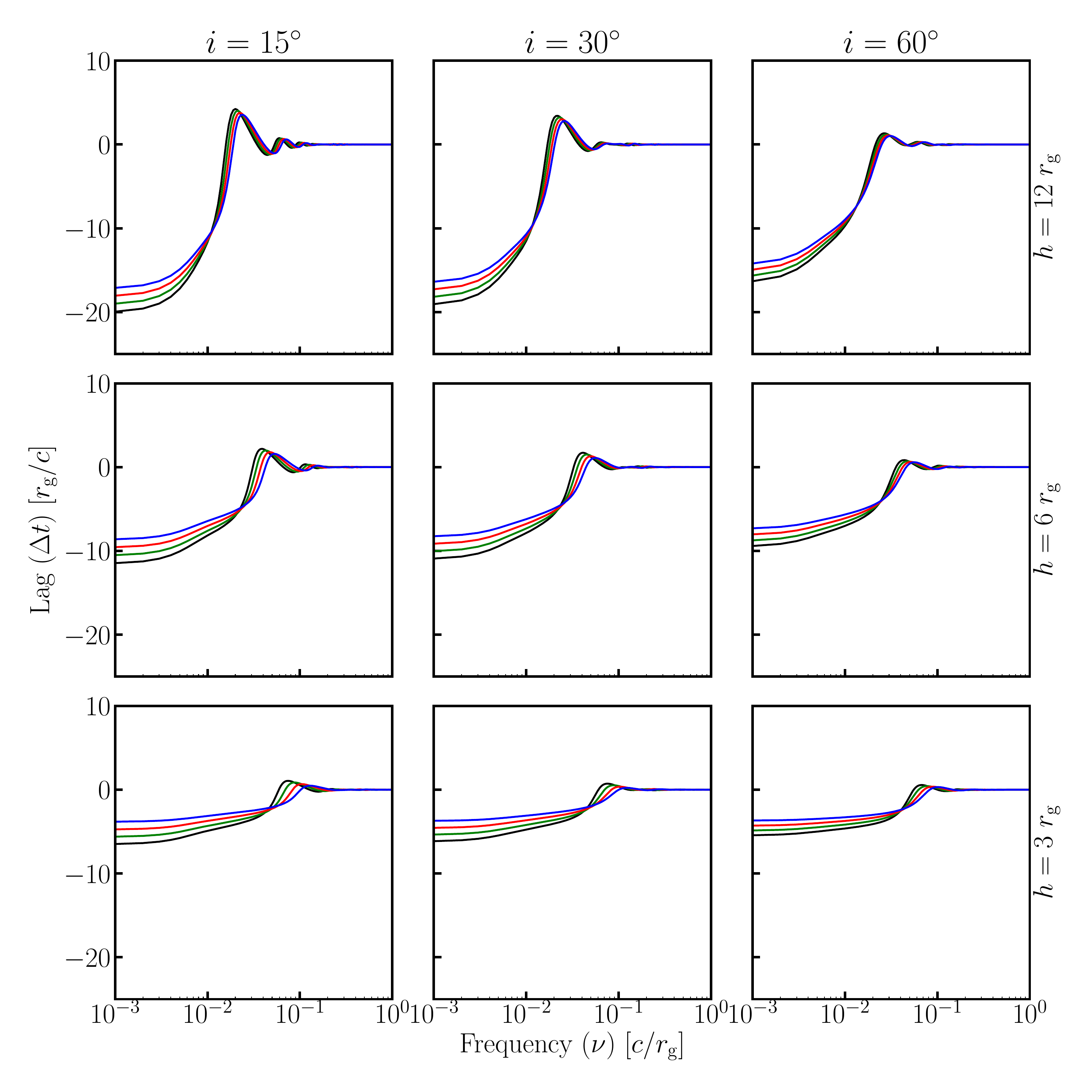}
\caption{Same as Figure \ref{lag-freq1}, but with a rapidly-spinning black hole ($a$ = 0.99). The qualitative effects that disk thickness has on the lag-frequency spectrum are consistent with those presented in Figure \ref{lag-freq2}.} 
\label{lag-freq3}
\end{figure*}

Figures \ref{lag-freq1}, \ref{lag-freq2}, and \ref{lag-freq3} present the lag-frequency spectra for the cases of a black hole at $a$ = 0.00, 0.90, and 0.99 respectively, using a finite thickness disk with $\dot{M}$ = 0.1, 0.2, and 0.3 $\dot{M}_{\rm Edd}$ (grey, red, and blue respectively) as well as a razor-thin disk (black). From Figure \ref{lag-freq1} one sees that increasing disk thickness results in a decrease in the lag at $\nu$ $<$ $10^{-2}$ $c$/$r_{\rm g}$, as well as slightly increasing the phase wrapping frequency, all consistent with the decrease in the path difference between direct and reprocessed flux when increasing disk scale height. At this particular spin value, one also sees a steepening of the slope of the lag-frequency spectrum when $h$ = 3 and 6 $r_{\rm g}$, while the magnitude of the positive phase wrapping peak seems to be positively correlated with $\dot{M}$ when $i$ = 60\degree or when $i$ = 15\degree and $h$ = 3 or 6 $r_{\rm g}$, while the opposite is the case when $h$ = 12 $r_{\rm g}$ and $i$ = 30\degree.

The negative correlation between the reflection lag magnitude and disk thickness is also seen when $a$ = 0.90 and 0.99 in Figures \ref{lag-freq2} \& \ref{lag-freq3}, though the change in the magnitude is not as dramatic as the Schwarzschild case. This is due to the decrease in disk thickness with increasing spin, as $z$ is inversely proportional to efficiency radiative efficiency $\eta$, which itself is positively correlated to $a$ as $\eta$ is commonly defined (see Section \ref{sec:methods}). Once again, one sees an increase in the phase wrapping frequency, and in all cases, the magnitude of the first positive phase wrapping peak is negatively correlated to $\dot{M}$. From these results, together with those presented in the previous figure, one would expect that the use of the razor-thin disk approximation when modeling data taken from sources with non-negligible disk vertical structure would result in an underestimation of the coronal height.

\begin{figure*}
\centering
\includegraphics[width=0.96\linewidth]{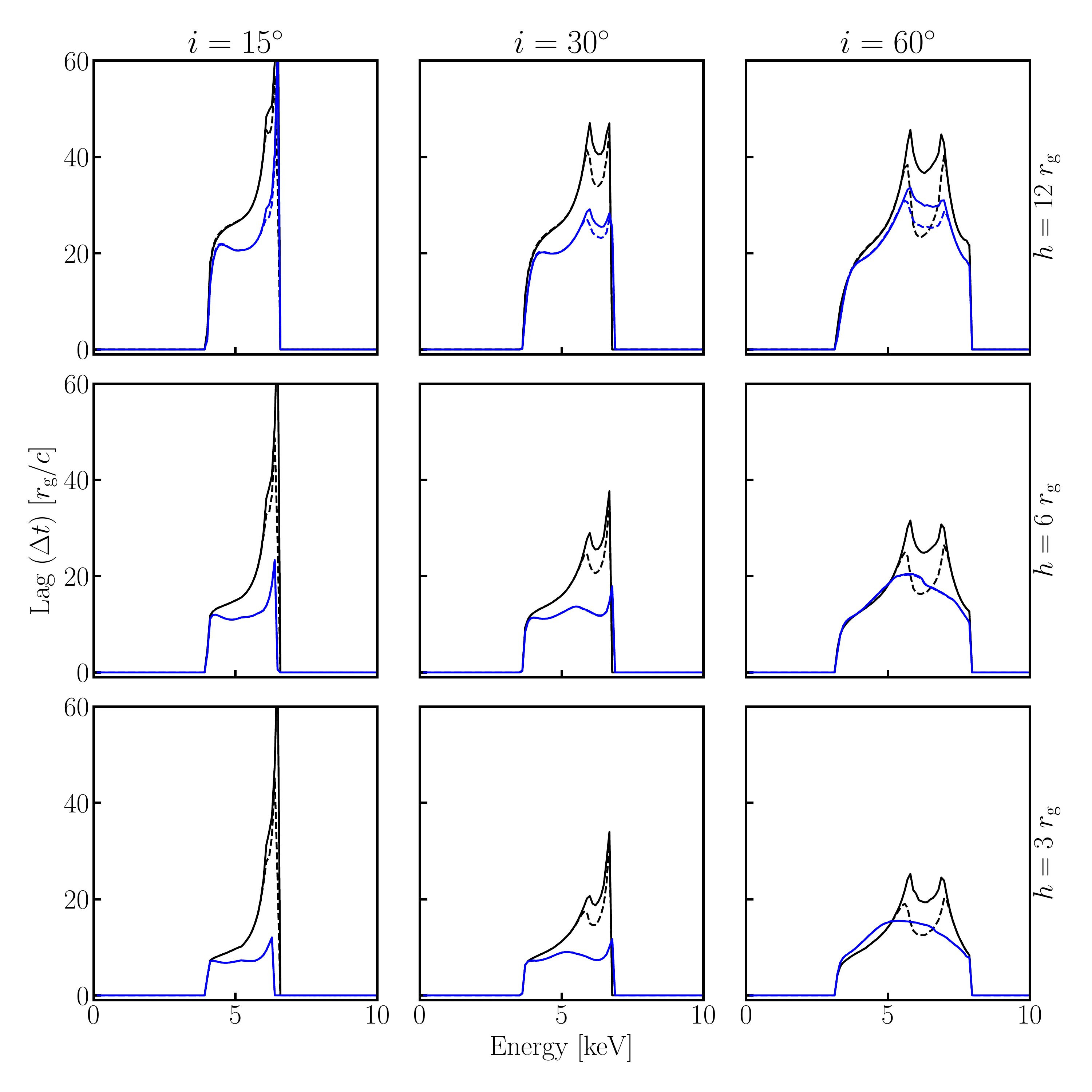}
\caption{The lag-energy spectra $[\Delta t(E)]$ for a Schwarzschild black hole ($a$ = 0.00) and has either a razor-thin disk (black) or a disk with a half-thickness given by Equation \ref{eq:ss73} with an associated mass accretion rate $\dot{M}$ = 0.3 (blue) $\dot{M}_{\rm Edd}$, the disk assumed to be neutral [$E_{\rm rest}(Fe \, K\alpha)$ = 6.4 keV]. These lag-energy spectra were created by averaging the energy-dependent lag over the frequency ranges $\nu$ = $1-3\times10^{-3} \, c/r_{\rm g}$ (solid line) and $5-8\times10^{-3} \, c/r_{\rm g}$ (dashed line), and are shown for the model parameters from Figure \ref{lag-freq1}. As the disk thickness increases, the average lag likewise decreases in all energy bins, with the one exception being presented in the case of $h$ = 3 $r_{\rm g}$ and $i$ = 60\degree (bottom right). When the lamppost corona is close to the black hole ($h$ = 3 $r_{\rm g}$, 6 $r_{\rm g}$), the low and high frequency lag-energy spectra are nearly indistinguishable, consistent with the broadening of the signal in the frequency domain due to the suppression of the late-time signal in the time-domain by self-shielding.} 
\label{lag-energy1}
\end{figure*}

\begin{figure*}
\centering
\includegraphics[width=0.96\linewidth]{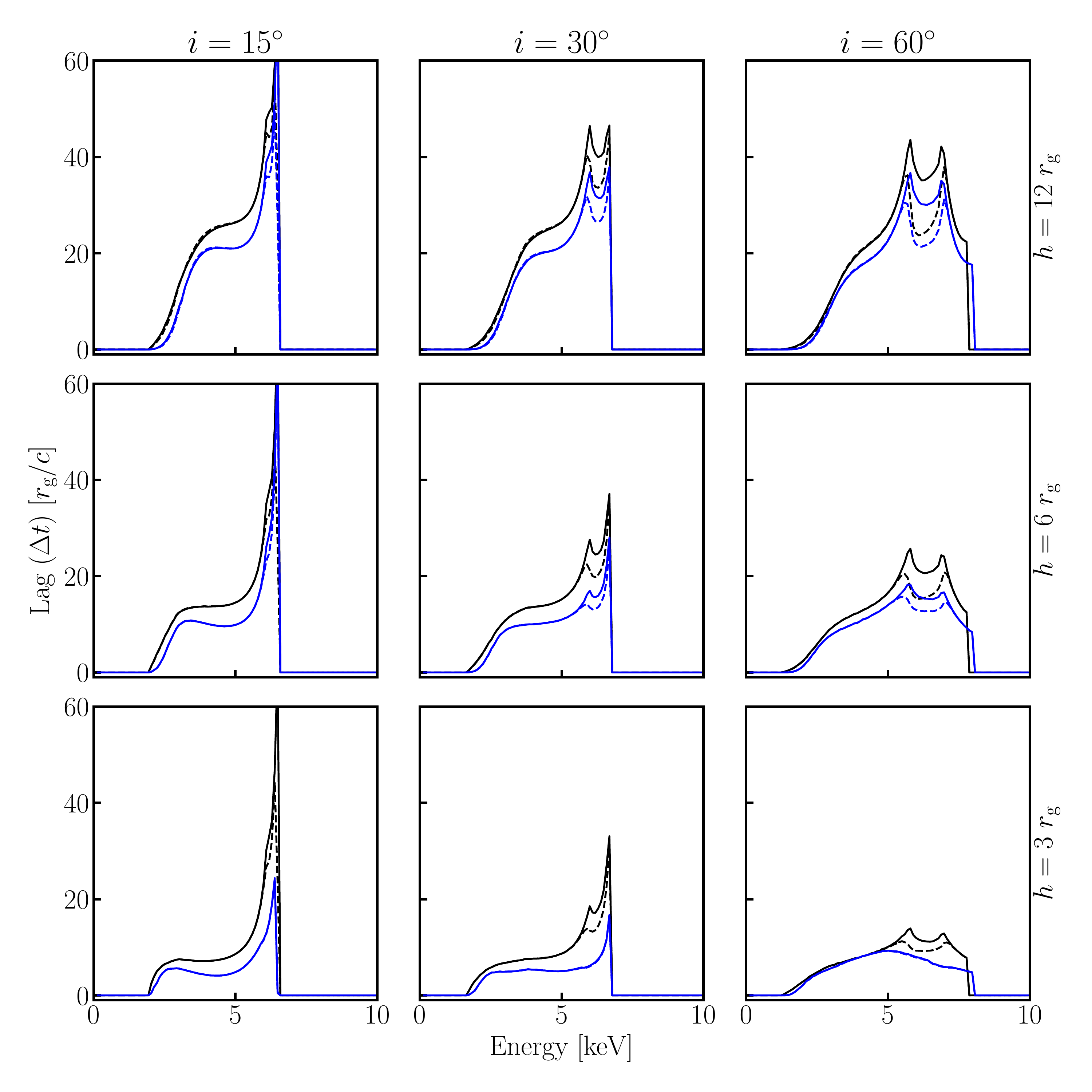}
\caption{Same as Figure \ref{lag-energy1}, but with a spinning black hole at $a$ = 0.9. Once again, an increase in disk thickness will typically result in the decrease in the average lag. Comparing the change in lag magnitude with Figure \ref{lag-energy1}, one finds the decrease is not as drastic as in the Schwarzschild case, consistent with a thinner disk due to increase radiative efficiency ($\eta$). Furthermore, one can distinguish between the low and high frequency lag-energy spectra for $h$ = 6 $r_{\rm g}$ as the decrease in disk thickness has lessened the effects of self-shielding at this particular corona height.}
\label{lag-energy2}
\end{figure*}

\begin{figure*}
\centering
\includegraphics[width=0.96\linewidth]{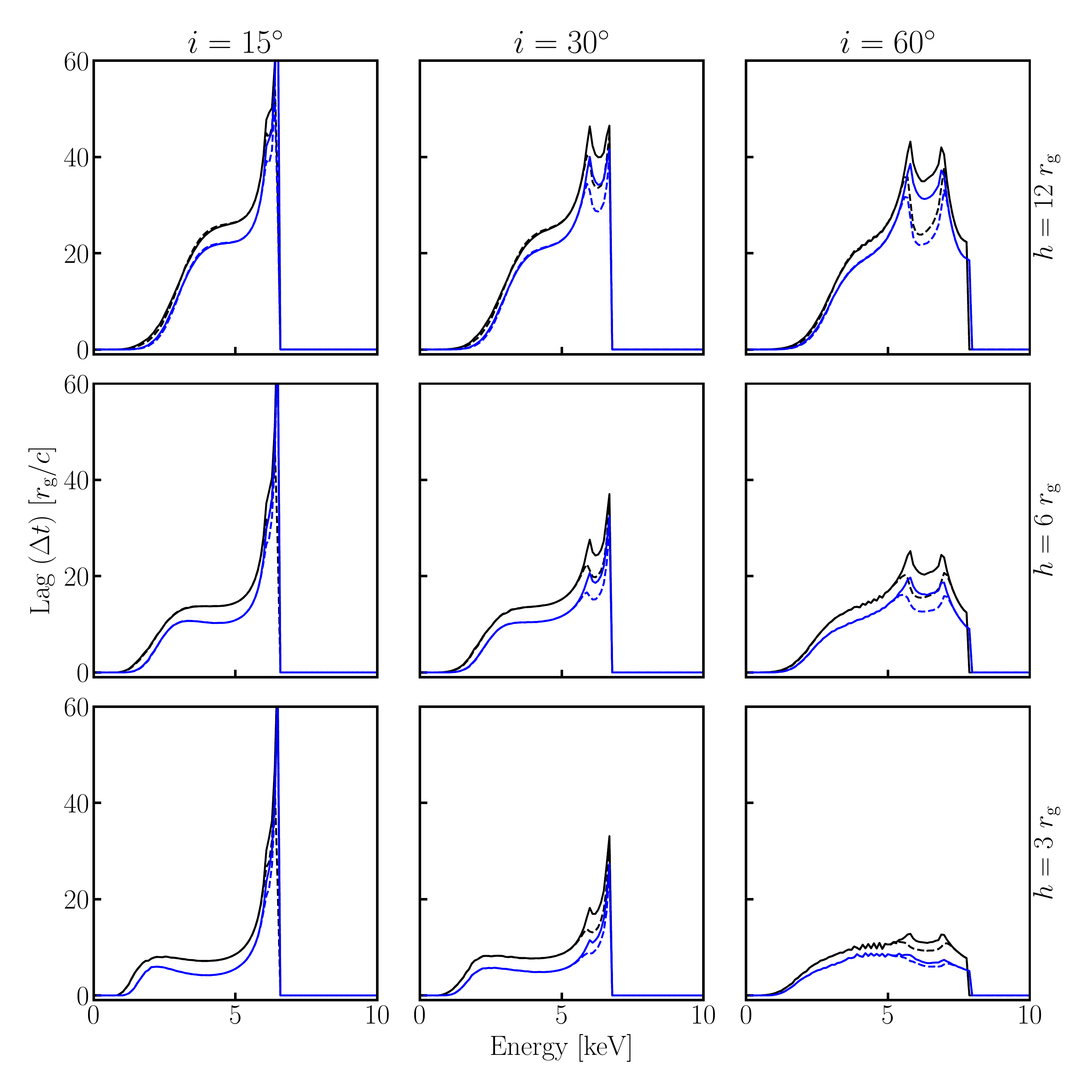}
\caption{Same as Figure \ref{lag-energy2}, but with a rapidly-spinning black hole at $a$ = 0.99. As expected, the decrease in the lag magnitude is not as dramatic compared to Figures \ref{lag-energy1} and \ref{lag-energy2}.} 
\label{lag-energy3}
\end{figure*}

Figures \ref{lag-energy1},\ref{lag-energy2}, and \ref{lag-energy3} present the lag-energy spectrum for the values of ($a$,$h$,$i$) presented in the previous three figures, where we include the spectra for a razor-thin disk (black) and a finite-thickness disk at $\dot{M}$ = 0.3 $\dot{M}_{\rm Edd}$ (blue), averaged over the frequency ranges $\nu$ = $1-3\times10^{-3} \, c/r_{\rm g}$ (solid line) and $5-8\times10^{-3} \, c/r_{\rm g}$ (dashed line). Consistent with the lag-frequency spectra, one finds that the lag-magnitude is typically decreased in the finite-thickness case (e.g. lower right of Figure \ref{lag-energy1}) as compared to its razor-thin counterpart. This trend is not true at all energies for all values of $i$, with the lag slightly increasing at lower energies for the case of ($a$,$h$,$i$) = (0.0, 3 $r_{\rm g}$, 60\degree) (Figure \ref{lag-energy1}) and at high energies for $a$ = 0.90 and $i$ = 60\degree (Figure \ref{lag-energy2}). A final notable effect is that, for the case of $\dot{M}$ = 0.3 $\dot{M}_{\rm Edd}$ when the corona is very close to the event horizon (at $h$ = 3 $r_{\rm g}$), one finds that the low and high frequency lag-energy spectra overlap are often almost indistinguishable from each other. This was seen previously in Figures \ref{transfers1} and \ref{transfers2}, and is naturally explained by the suppression of the late-time response from the outer disk by self-shielding, thus narrowing the signal in the time domain and broadening the signal in the frequency domain. Comparing the change in magnitude across the different values of spin, one finds that the suppression of lag magnitude is roughly inversely correlated with $a$, consistent with the decrease in disk thickness due to the increase in efficiency $\eta$ with increasing spin.

\subsection{The Disk-Hugging Corona}\label{sec:offaxis}

For exploring the case of an off-axis corona, we chosen to position the corona at a cylindrical radius $\rho_{\rm c} \in \{r_{\rm ISCO}, 1.5\,r_{\rm ISCO}, 2\,r_{\rm ISCO}, 2.5\,r_{\rm ISCO}\}$, a height $h_{\rm c} \in \{0.1\,r_{\rm g}, 0.5\,r_{\rm g}\}$ above the surface of the disk, and an azimuthal angle $\phi_{\rm c}$ between $0\degree$ and $360\degree$ in increments of 10\degree. We performed these calculations for spins of $a$ = 0.00 and 0.90 using a finite thickness disk with $\dot{M}$ = 0.3 $\dot{M}_{\rm Edd}$, as well as a razor-thin disk as a control. It is important to note that, as $h_{\rm c}$ is measured from the disk surface, the distance of the corona above the disk mid-plane will vary with disk thickness for any given value of $h_{\rm c}$.

\begin{figure*}
\centering
\includegraphics[width=0.49\linewidth, height = 0.42\linewidth]{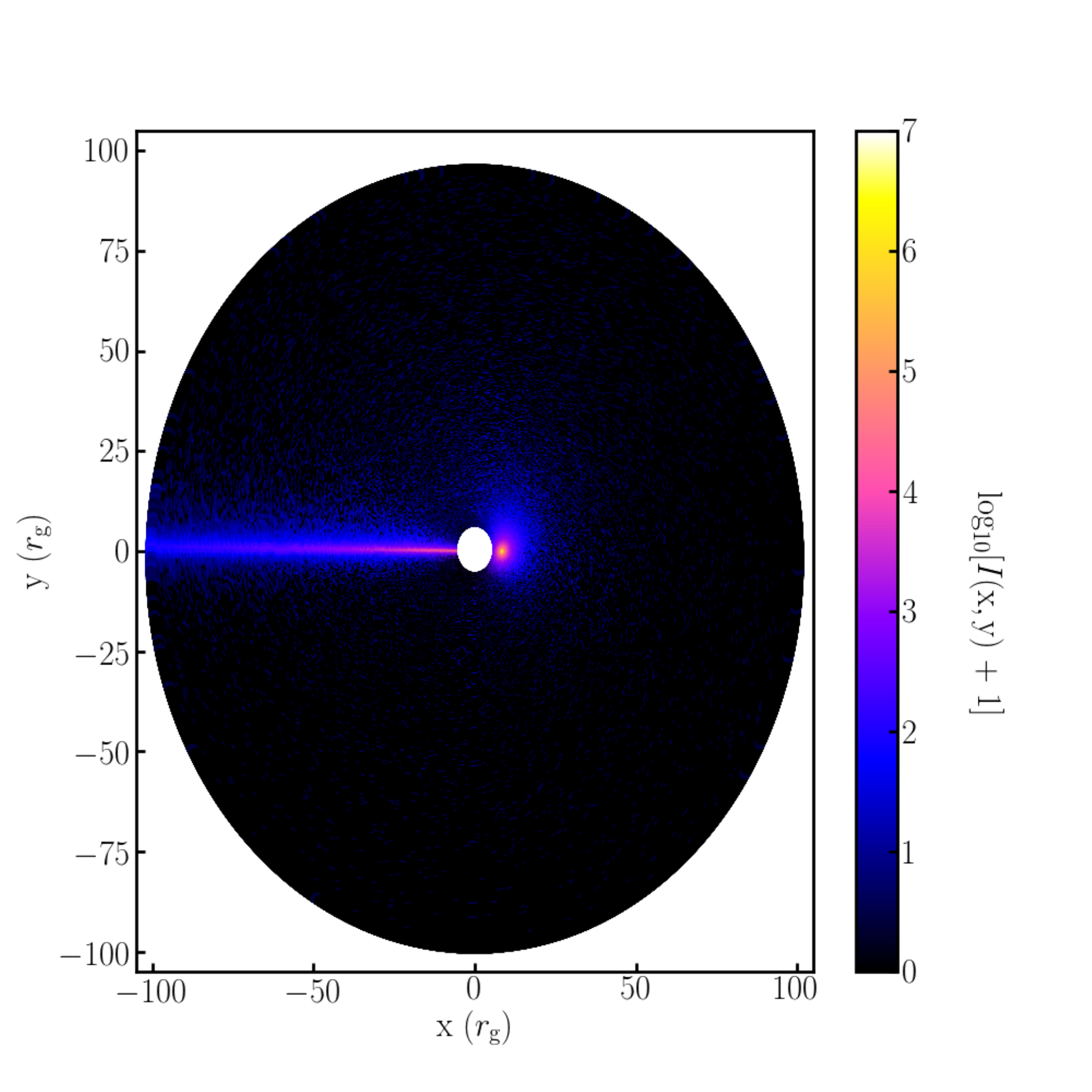}
\includegraphics[width=0.49\linewidth, height = 0.42\linewidth]{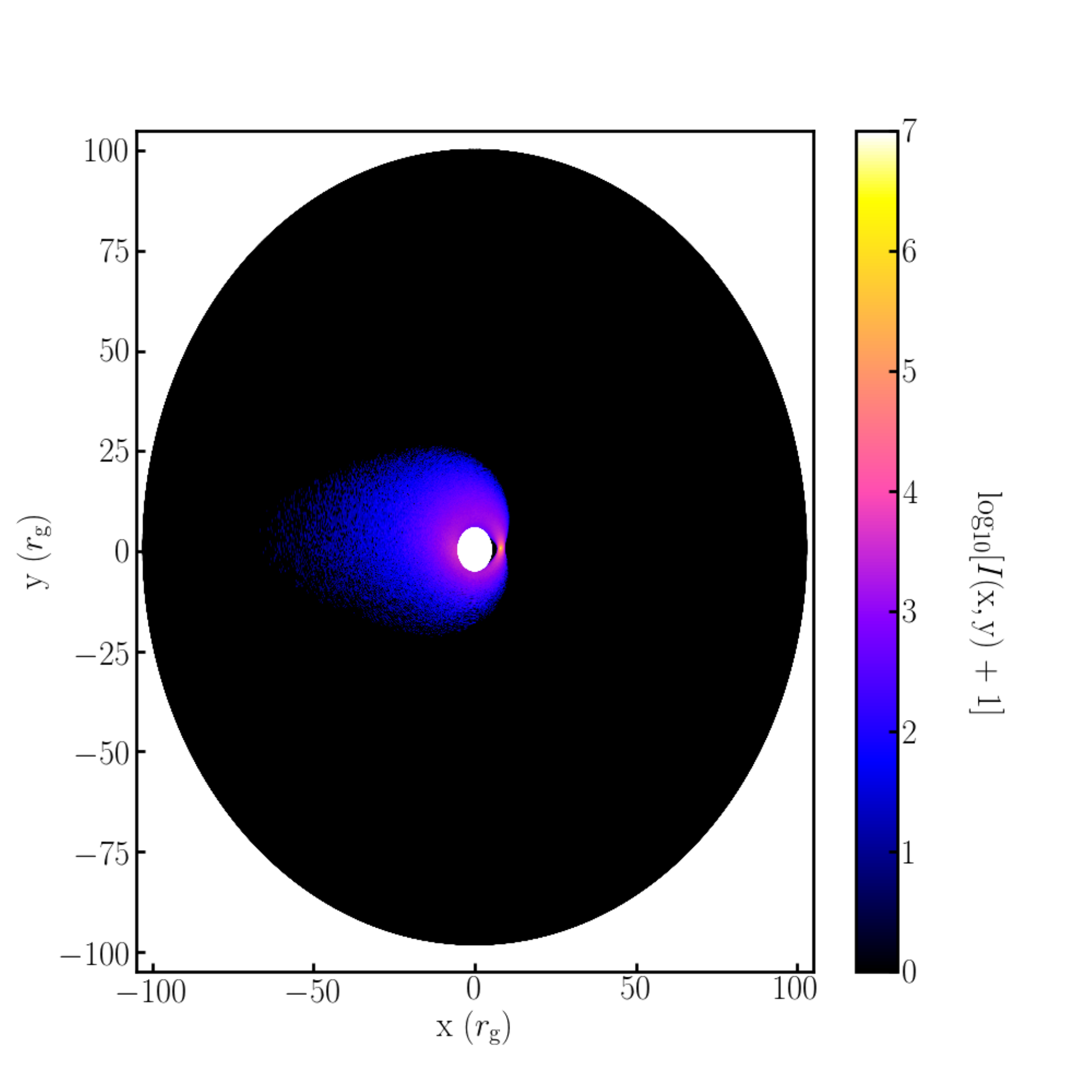}
\includegraphics[width=0.49\linewidth, height = 0.42\linewidth]{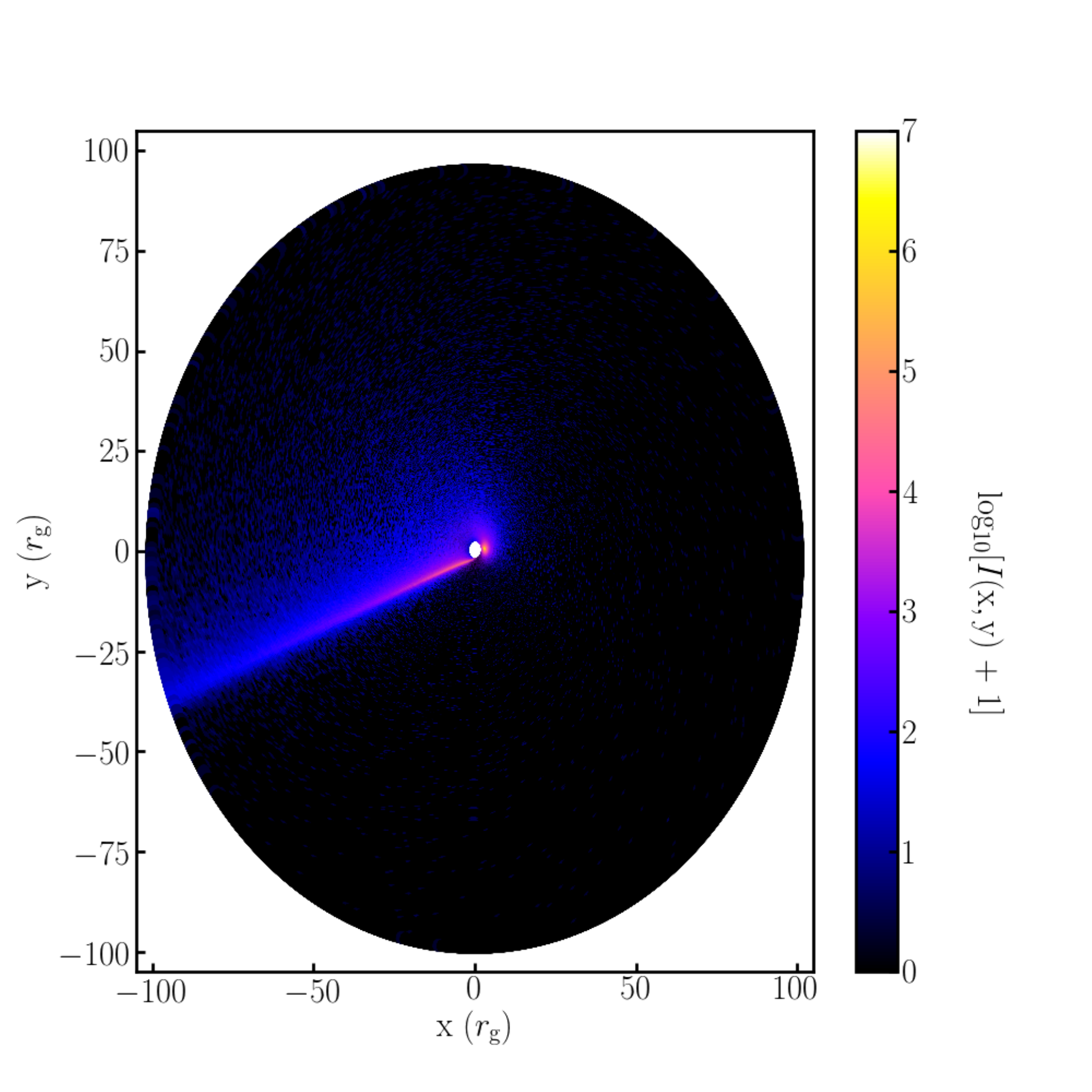}
\includegraphics[width=0.49\linewidth, height = 0.42\linewidth]{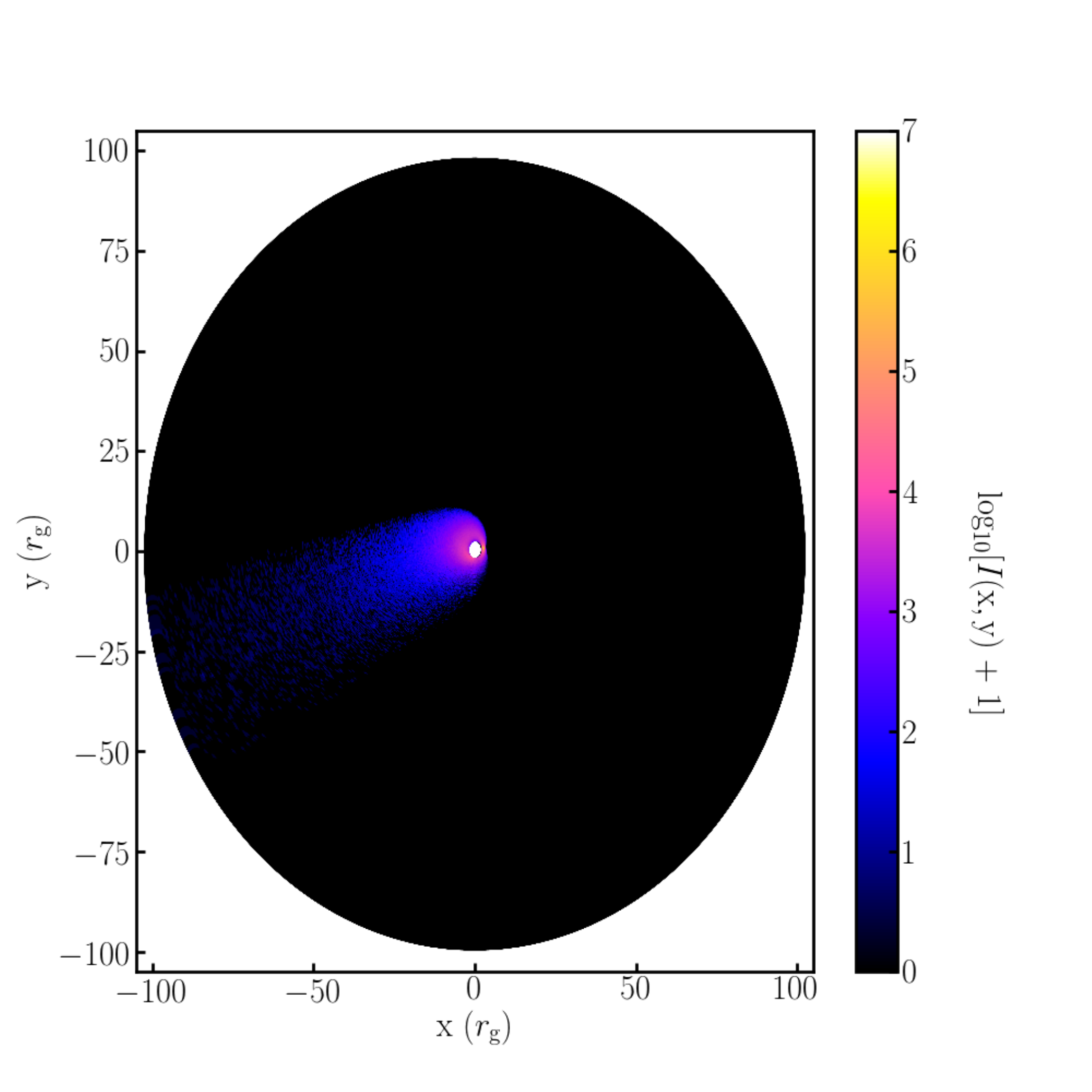}
\caption{Reflection intensity maps for an off-axis corona orbiting around a Schwarzschild ($a$ = 0.00, top) and a spinning ($a$ = 0.90, bottom) black hole. In both cases, the corona is situated at a cylindrical radius of $\rho_{\rm c}$ = 1.5 $r_{\rm ISCO}$ and an azimuthal angle of $\phi_{\rm c}$ = 90\degree relative to the observers line of sight, seen at $i$ = 15\degree. The left column is the case of a razor-thin accretion disk, while the right is that of a finite thickness disk with $\dot{M}$ = 0.3 $\dot{M}_{\rm Edd}$, with the corona situated at a height $h_{\rm c}$ = 0.1 $r_{\rm g}$ above the surface of the disk in each case and orbiting with the disk. As one would expect, there is a small high-intensity patch on the disk right under the position of the corona in both cases, with a secondary patch of high intensity on the opposite side of the black hole, whose exact $\phi$ is a function of $a$ due to frame dragging. At the razor-thin limit, there appears to be a very narrow locus of points that constitute this secondary patch, explained naturally as a result of gravitational optics. The observed secondary patch gives some credence to our hypothesis of "cross-bowl" reverberation, where an off-axis corona configuration could produce a lag signature from irradiating the accretion disk on the opposite side of the black hole.} 
\label{example-offaxis_disks}
\end{figure*}

In Figure \ref{example-offaxis_disks}, we show the reflection intensity maps, that is intensity as a function of position of the disk image seen by the observer, for an off-axis corona (say, a magnetic reconnection event close to the surface of the disk) around Schwarzschild ($a$ = 0.00, top) and spinning ($a$ = 0.90) black holes,  presented with both a razor-thin (left) and finite-thickness (right) disk. The corona has been given the parameters of ($\rho_{\rm c}$, $h_{\rm c}$, $\phi_{\rm c}$) = (1.5 $r_{\rm ISCO}$, 0.1 $r_{\rm g}$, 90\degree) in each case, and as one expects, there exists a small high-intensity patch roughly at the position of the corona on the intensity map (we will call this the "primary" patch), consistent with a strong irradiation of portion of the disk directly under the corona. In all cases, there is a secondary intensity patch on the side of the disk opposite that of the primary patch, the $\phi$ position of which varying with $a$ due to frame dragging, suggesting that an off-axis corona can irradiate the disk "across the bowl" in both disk geometries. The fact that the secondary patch is such a narrow locus of points in the razor-thin limit in the Schwarzschild geometry is a natural consequence of photons being confined to the orbital plane defined by their origin, the black hole singularity, and their initial momentum vector. Such photons can only ever return to the (razor-thin) disk at an azimuth $\phi$ = $\phi_{c}$+180\degree. This symmetry is broken by frame dragging when $a$ $>$ 0.00, resulting in the secondary response being at $\phi$ $>$ $\phi_{\rm c}$ + $180\degree$ and a slight broadening of the $\phi$ distribution due to the photons being dragged in the positive-$\phi$ direction (black hole is spinning counter-clockwise).

This cross-bowl irradiation results in a natural lag due to path length difference between the direct corona light (which would have a very similar arrival time as that of the reprocessed radiation from the primary patch) and the radiation coming from this secondary patch. As such, it is interesting to consider whether we may mimic the reverberation characteristics of a lamp-post corona with a corona that is "hugging" a thickened disk, and it is this that we will focus on in this section.

\begin{figure*}
\centering
\includegraphics[width=0.48\linewidth]{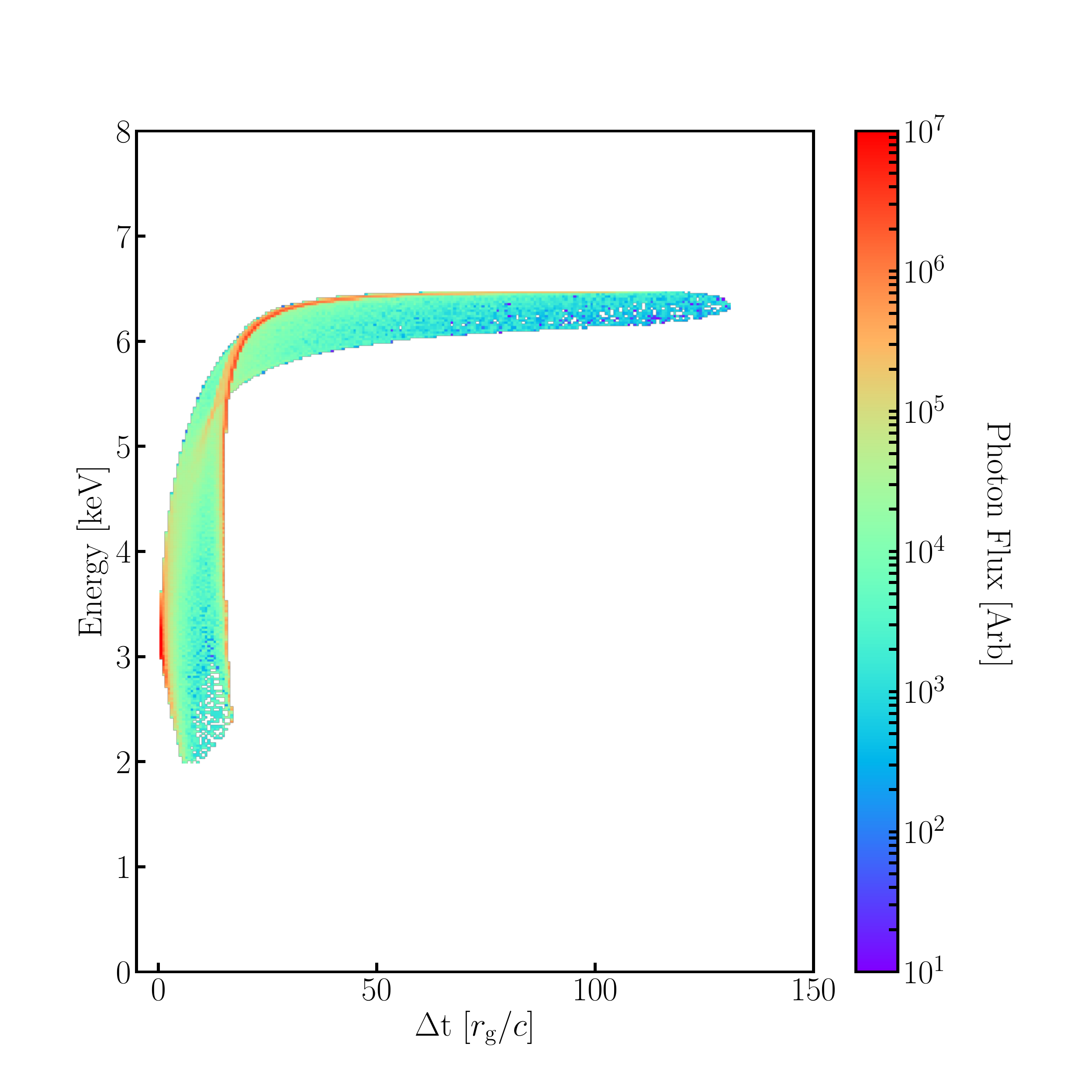}
\includegraphics[width=0.48\linewidth]{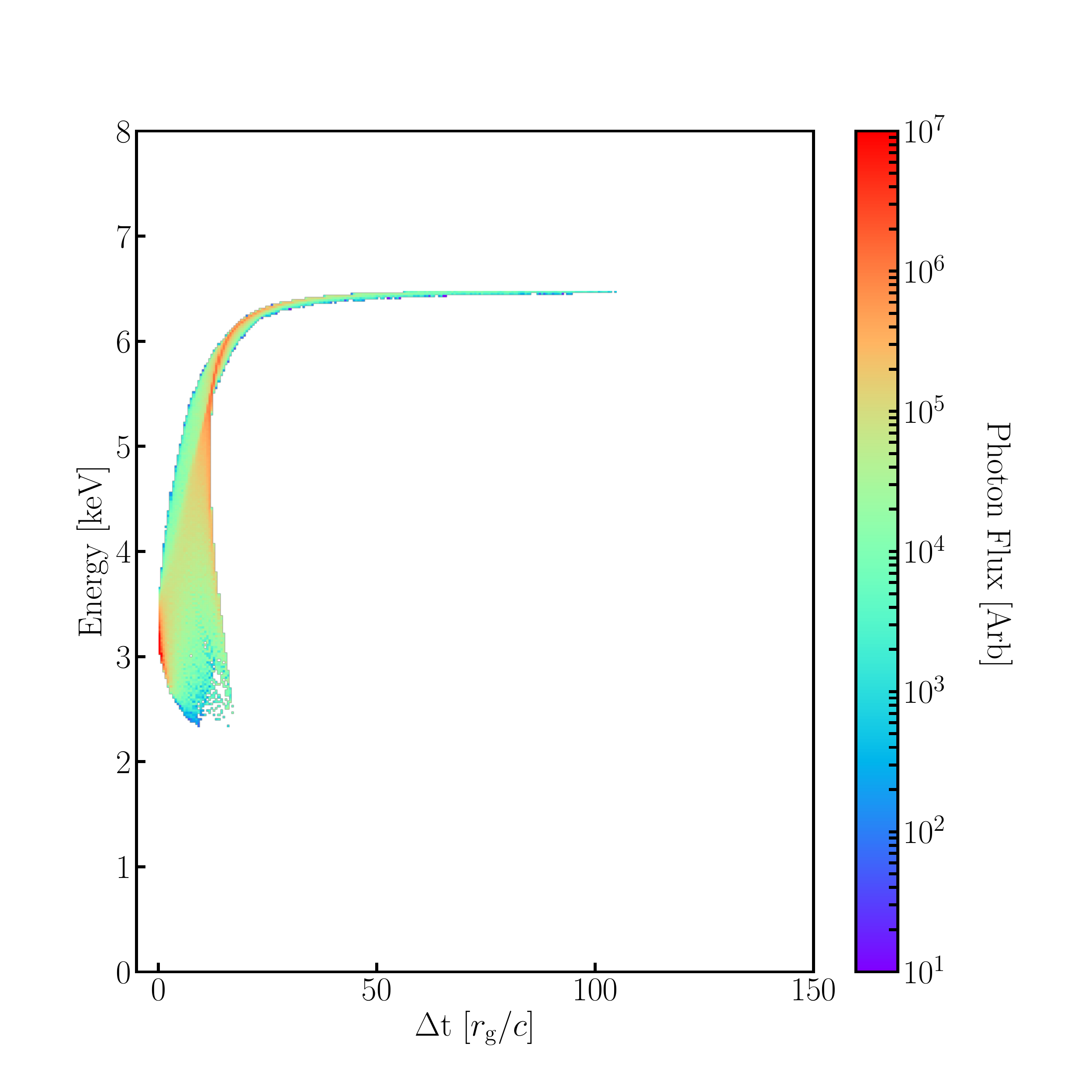}
\caption{The transfer functions for the Fe K$\alpha$ response from the off-axis coronae presented in the bottom row of Figure \ref{example-offaxis_disks}, where we have once again presented the razor-thin disk case (left) and the finite thin disk case with $\dot{M}$ = 0.3 $\dot{M}_{\rm Edd}$ (right). There are two strong peaks in the photon flux near $\Delta t$ $\sim$ 0 and 20 $r_{\rm g}/c$, being from the primary and secondary high-intensity regions presented in the previous figure, each appearing several orders of magnitude greater than the rest of the transfer function. It is most likely that these high-intensity patches would dominate the signal, and thus any late-time signal would be likely undetectable in flux-limited studies of AGN.}
\label{example-offaxis_transfers}
\end{figure*}

Figure \ref{example-offaxis_transfers} presents the transfer functions associated with the corona position and disk thickness presented in the bottom row of Figure \ref{example-offaxis_disks}, once again with the razor-thin disk being in the left panel and the finite thickness case in the right panel. As expected from the intensity maps, these transfer functions both double-peaked, with the primary and secondary responses being at $\Delta t$ $\sim$ 0 and $\sim \, 20 \, r_{\rm g}/c$ respectively. We find that, in general, this double-peak is ubiquitous in our 2D transfer functions for off-axis coronae. Looking at the color scale, it is immediately clear that there is a very large dynamic range, with the primary and secondary responses having photon fluxes that are several orders of magnitude larger than the other portions of the transfer function, and thus qualitatively inconsistent with the lamppost corona model which has its flux more distributed throughout the E-$\Delta t$ space.

However, if a black hole had such a disk-hugging corona above an axisymmetric accretion flow, there is no reason to assume such a corona would have a favored value of $\phi_{\rm c}$, and thus for a given spin and inclination angle, a transfer function corresponding to a single choice of the ordered triple ($\rho_{\rm c}$, $h_{\rm c}$, $\phi_{\rm c}$) would be inadequate to describe the reverberation characteristics observed over an entire light curve. As such we created effective 2D transfer functions by taking the mean of the individual un-normalized 2D transfer functions across all values of $\phi_{\rm c}$. For a set of $N$ transfer functions $\{\Psi_{\rm j} | j = 1, 2, 3, ..., N\}$, each $j$ corresponding to a separate $\phi_{\rm{c},j}$,

\begin{equation}
\centering
\Psi_{\rm eff}(E,t) = \frac{1}{N}\sum_{j = 1}^{N}\Psi_{j}(E,t)
\label{eq:effective-transfer}
\end{equation}
The choice of taking the average of the un-normalized transfer functions was done as a natural way to weigh each case by its photon flux, as each case has the same luminosity in the co-moving coronal frame and thus their observed un-normalized photon fluxes can be directly compared. As noted in Section \ref{sec:methods}, if one assumes that the lag signal is purely coherent, the full light curve can be modeled as a series of individual flash events and their corresponding disk responses, with Equation \ref{eq:lightcurve1} being rewritten as a series. This makes this form of the effective transfer function quite natural, and given the linear nature of the Fourier transform, one can generate the appropriate cross-spectrum (and corresponding lag) by substituting $\Psi_{\rm eff}$ into Equations \ref{eq:cross-spec} and \ref{eq:lag}, especially as the reflection fraction $R$ would not vary with $\phi_{\rm c}$ given axisymmetry.

\begin{figure*}
\centering
\includegraphics[width=0.48\linewidth]{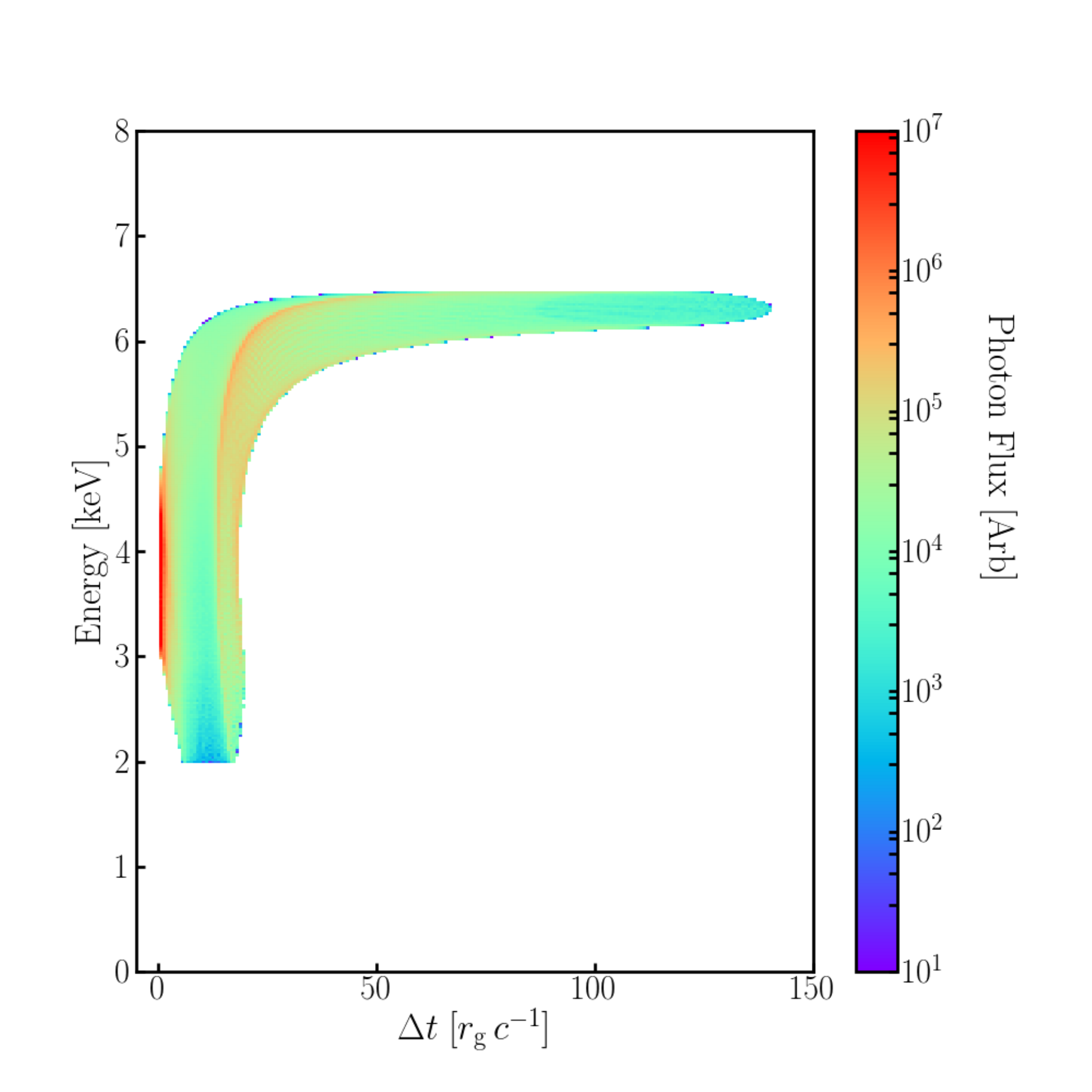}
\includegraphics[width=0.48\linewidth]{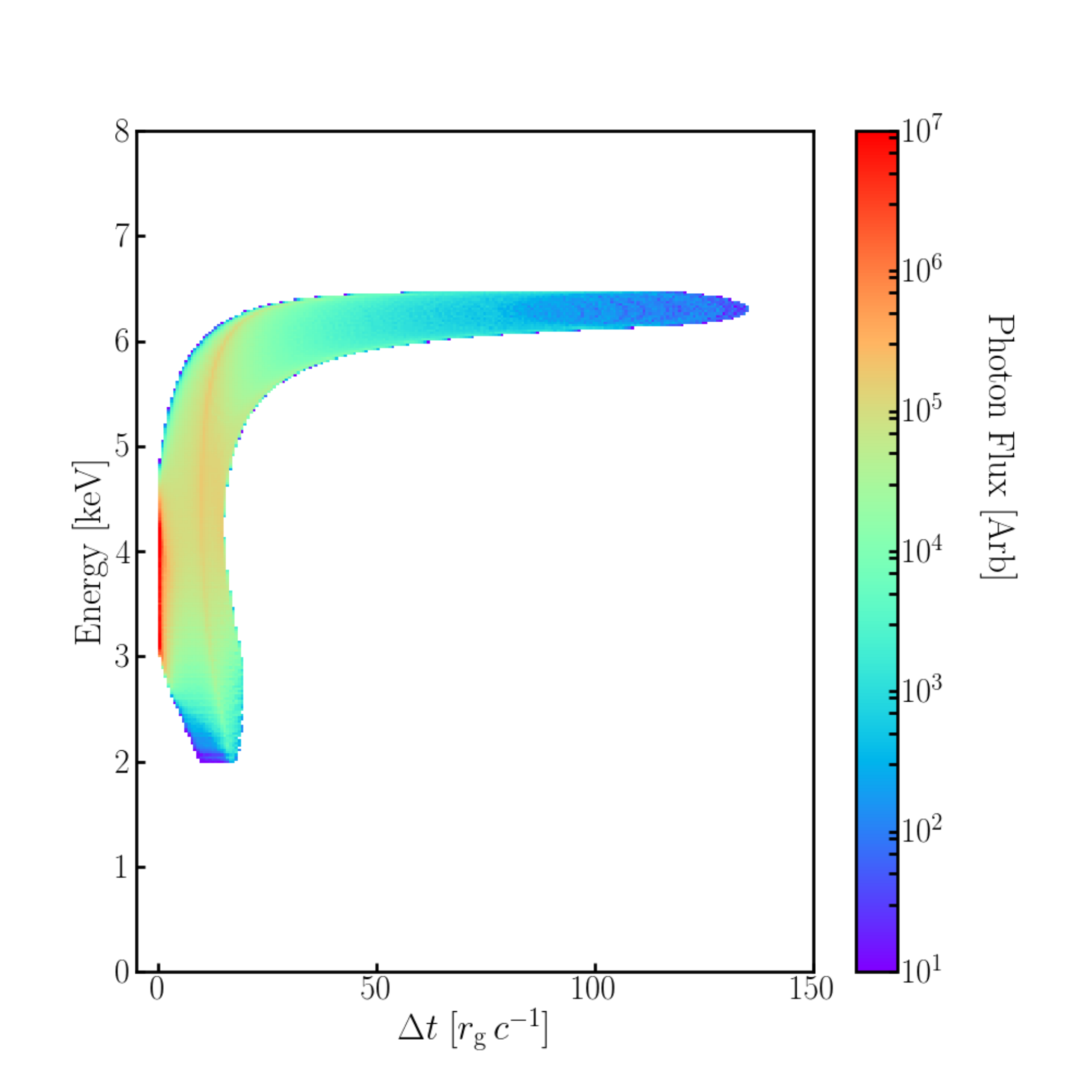}
\caption{Effective transfer functions created by taking the numerical average of the off-axis transfer functions for a series of $\phi_{\rm c}$ between 0\degree and 350\degree inclusive, in increments of $\Delta \phi_{\rm c}$ = 10\degree, and all with $\rho_{\rm c}$ = 1.5 $r_{\rm ISCO}$ and $h_{\rm c}$ = 0.1 $r_{\rm g}$. This is for the same $a$ and disk thicknesses presented in Figure \ref{example-offaxis_transfers}. One sees once again that the line response is dominated by primary and secondary regions, albeit with over a greater spread of photon energies compared to the single $\phi_{\rm c}$ case presented previously. While reverberation "across the bowl" appears to happen in both disk geometries, the primary response region has a flux that is approximately 95\% greater than the secondary region, and thus would dilute the average lag signal. This is inconsistent with the moderate lags seen in AGN lag-frequency spectra, and thus suggests that a "disk-hugging" corona is inadequate to explain the high-frequency soft lags, and reinforces the common assumption that the irradiating corona must be physically separated from the reprocessing material.} 
\label{example-multiphi_transfers}
\end{figure*}

Figure \ref{example-multiphi_transfers} give two examples of $\Psi_{\rm eff}(E,t)$ for the case of a spinning black hole ($a$ = 0.9), observed at an angle $i$ = 60\degree, the irradiating annulus being at a cylindrical radius of $\rho_{\rm c}$ = 1.5 $r_{\rm ISCO}$ and $h_{\rm c}$ = 0.1 $r_{\rm g}$ above the surfaces of a razor-thin (left) and finite thickness disk (right). Looking at these two examples, while one sees much more extended transfer functions than in the previous case, the primary and secondary response patches still dominate the signal. It is also apparent that there is a large difference in the photon flux between the primary and secondary patches, with the secondary having $\leq 5\%$ the photon flux of the primary. Given that the primary reflection response has a negligible lag magnitude, this would naturally dilute any potential lag signature quite dramatically, which then would be further diluted by the direct coronal flux. One possible way to get around this would be if the region directly underneath the corona would be irradiated to the point of complete ionization, however this would require a very strong coronal event, the lag of which would be naturally $\Delta t$ = 0, and thus again likely diluting the lag signal to below the point of detection. 

We can conclude that, while "cross-bowl" disk irradiation is possible, this alone would not be sufficient to explain the observed high-frequency lags in AGN if the corona is "disk-hugging". This is consistent with the common view that the corona responsible for the high frequency lag characteristics, such as the Fe K$\alpha$ reverberation, must be physically separate from the reprocessing material \citep{Reis+Miller2013}.

\section{Discussion} \label{sec:discussion}

As shown in the previous section, disk thickness can have dramatic effects on the 2D reverberation transfer function, imprinting itself on the lag-frequency and lag-energy spectra of AGN. Assuming a lamp-post corona configuration and a scattering surface consistent with that of \cite{Shakura+Sunyaev1973}, one finds that self-shielding is the predominant effect in the area of parameter space this work explored, where the convex geometry of the inner disk acts as a barrier for coronal photons, resulting in the suppression of the irradiation profile (and thus the emissivity profile) at larger radii; this effect was first reported in our earlier work \cite{Taylor+Reynolds2018} in the context of the time-average reflection spectrum. This results in the suppression of the late-time "blue-wing" of the transfer function when $h$ is small and a "hollowing" when $i$ is large (i.e. when the disk is seen more edge-on). When $h$ is larger (e.g. 12 $r_{\rm g}$), while the blue-wing is no longer suppressed, one finds a change in the slope of the "red-wing" due to a decrease in the delay between the observed corona flash and the initial response from the disk.

This suppression of late-time signatures results in an overall decrease in the lag magnitude of the lag-frequency spectrum with increasing disk thickness, the spread of the change of the lag decreasing with increasing black hole spin due to a inverse relationship between radiative efficiency $\eta$ and disk thickness. This suppression of the lag magnitude is also ubiquitous in the lag-energy spectrum, along with a suppression of frequency dependance to the lag-energy profile. In the context of observation, for a fixed reflection fraction, this overall decrease in the high-frequency lag signal would naturally cause an underestimation in the distance between the corona and the reprocessing material (or in the context of the lamp-post assumption, the coronal height $h$). For example, if we assume we knew $a$ and $i$ for the case presented in Figure \ref{transfers3}, the change in the lag-frequency spectrum ($\sim$ 7 $r_{\rm g}$/$c$ at $\nu$ = $10^{-3}$ $c$/$r_{\rm g}$) would likely result in a best fit height of $h$ $\sim$ 5 $r_{\rm g}$. This is interesting given that there is an abundance of small coronal heights quoted as best-fit values in the literature (e.g. \citealt{Dauser+2012, Kara+2015, Kara+2016, Frederick+2018}), however it must be noted that such an effect can also be achieved by decreasing the reflection fraction by approximately $\sim$ 33\% and thus diluting the lag signal by approximately the same amount \citep{Uttley+2014}. Thus, fitting the lag-frequency spectrum with both a free reflection faction and $\dot{M}$ is likely to result in further degeneracies in $\chi^{2}$ space. Ultimately, progress must be made by simultaneously using spectral and reverberation timing data to break these degeneracies.

Finally, we have explored the scenario of an off-axis "disk-hugging" corona (absent a lamp-post), asking if such a corona could mimic a lamp-post-like signal, using both single point source flashes and an extended annulus. While irradiation by returning radiation from such a corona does spark curiosity, we find that any potential lag signal is likely to be diluted beyond detectability, as $\sim 95\%$ of the observed flux would have negligible lag times, coming from the disk immediately underneath the corona. This dilution would be further enhanced by the continuum itself, which has a null lag in the standard reflection paradigm (i.e. without incorporating low-frequency hard lags). Another point to note is that the transfer functions (see Figure \ref{example-multiphi_transfers}) suggest that this reflected flux would be primarily heavily red-shifted to below 5 keV, and thus inconsistent with observations of the time-averaged broad Fe K$\alpha$ line in Seyfert I galaxies, peaking towards the rest energy $\sim 6-7$ keV. Thus, while the exact coronal geometry is unknown and we cannot rule out the possibility of a corona that is partially extended over the disk, we can conclude that a significant portion of the irradiating flux must be coming from a source that "stands off" from the accretion disk, such as that of a lamp-post or an extended jet \citep{Zoghbi+2012, Kara+2016, Wilkins+2016}.

We emphasize that this exploration does not take into account all complexities of reverberation modeling, and instead should be thought of as a proof of principle rather than a true statement of reality. While we have included a more physically-motivated accretion disk model as compared to the razor-thin approximation, the exact geometry of the accretion disk remains open to debate. Also, as the \cite{Shakura+Sunyaev1973} thin disk approximation breaks down at roughly $\dot{M}$ $>$ 0.3 $\dot{M}_{\rm Edd}$, exploring reflection and reverberation in super-Eddington systems would require the implementation of alternative disk models, such as a scale height derived from an analytic accretion model \citep{Abramowicz+1980, Abramowicz+1988} or from the output of a GRMHD simulation \citep{McKinney+2014,Jiang+2014,Jiang+2017}. In the future, we hope to be able to expand beyond the simple thin disk assumption, and {\tt Fenrir} is already well-suited to do that, as stated in Section \ref{sec:methods}. In most of this work, we have also assumed a point-source corona that varies only by its magnitude, with its spectral index $\Gamma$ being constant. These are common assumptions in the literature, which we have chosen to use for consistency, but the inclusion of an extended corona would almost certainly add in more phenomenological complexities and is also a likely necessity for explaining the low-frequency hard lag \citep{Wilkins+2016,Chainakun+Young2017}, the exploration of which is beyond the scope of this work. As noted in \cite{Mastroserio+2018}, the inclusion of a pivoting power-law (i.e. $d\Gamma$/$dt$ $\neq$ 0) adds in non-linearities, and thus could confound the linearity implicitly assumed in much of X-ray reverberation methodology. Finally, we have not included ionization effects into our calculations, instead simply assuming a rest-energy of Fe K$\alpha$ is 6.4 keV. It is incredibly common to assume a single ionization state in reflection modeling (e.g. {\tt RELXILL} \citealt{Garcia+2014, Dauser+2014}), but this not likely to be true in AGN given that the incredibly centrally-concentrated emissivity profiles inferred from observation implies a radial ionization gradient across the disk \citep{Ross+Fabian1993, Reynolds+Fabian2008, Wilkins+Fabian2011}. While there have been recent attempts to incorporate such complexities into reverberation modeling, such as in {\tt KYNREFREV} \citep{Caballero-Garcia+2017, Caballero-Garcia+2018}, we have chosen to our simplified model for clarity. As this is the first exploration of the relationship between reverberation signatures and disk geometries, we have opted for clean interpretations by eliminating possible degeneracies that may arise, instead allowing more thorough explorations to be performed in the future.

\section{Summary} \label{sec:summary}

The study of X-ray reflection and reverberation in BHB and AGN has proven extremely fruitful, allowing us to gain a deeper understanding of the central black holes, as well as the plasma that resides in the hot electron corona and the accretion flow. Modeling reverberation is accomplished via raytracing, calculating the photon orbits in Kerr spacetime from the corona to the disk, then disk to observer. In performing these calculations, it is common to make simplifications, such as approximating the corona as a lamppost or assuming that the disk thickness is negligible (a "razor-thin" disk). Using {\tt Fenrir} \citep{Taylor+Reynolds2018}, we have explored the effects that disk thickness has on reverberation lags as a function of Fourier frequency and photon energy, approximating the disk as an optically thick, geometrically thick, radiatively dominated \cite{Shakura+Sunyaev1973} accretion disk, while still using the lamppost approximation for a fiducial model.

We found that the overall magnitude of the lag is consistently inversely correlated with disk thickness, with said change being inversely correlated with black hole angular momentum. This is apparent in both the lag-frequency and lag-energy spectra, with the decrease in the lag being greatest when the corona is close to the event horizon, where the inner edges of the disk act to prevent much of the flux from irradiating the outer regions of the disk. This "self-shielding" of the disk results in a truncating of the late-time "blue wing" of the 2D transfer function at all observer angles, while also "hollowing" out the transfer function at $i$ = 60\degree. Even with the corona well outside of the "bowl" of the disk, there is an overall decrease in the lag due to the decrease in the travel time between the corona and the disk, resulting in the changing of the slope of the relativistic "red wing" of the transfer function.

We can conclude that, for a given reflection fraction, a non-zero disk thickness would result in an underestimation of the lamppost height, and would result in natural degeneracies between reflection fraction and disk thickness if one were to analyze the lag-frequency spectrum in isolation. The effects of disk geometry on the lag-energy spectrum are qualitatively different from simple dilution however, with the shape of the relation being relatively unchanged as a function of Fourier frequency. We expect that these degeneracies to be overcome by analyzing the complex cross-spectrum as a unified unit, and such exploration is planned for the future.

Finally, we explored the possibility of off-axis "disk-hugging" corona, asking if such a source could produce similar reverberation characteristics as that of a lamppost, the lag signal being due to a coronal flash irradiating the side of the disk opposite itself. While reverberation "across the bowl" is possible, we find that such a scenario is inconsistent with observation. Most of the flux is observed to be coming from the region of the disk right under the corona (the primary region), with only a small fraction ($\sim$ 5\%) being seen coming from the secondary region opposite the primary. As the flux from the primary region would have a corresponding lag of $\sim$ 0, any potential lag signature would be diluted to the point of being undetectable. This is clearly inconsistent with the high-frequency soft lags observed in many Seyfert galaxies, thus requiring most of the irradiating to be coming from a source that is physically separated from the reprocessing material, such as a lamppost or an extended jet.
\newline
\newline
We would like to thank Cole Miller, Erin Kara, Dan Wilkins, Drew Hogg, Matt Middleton, Adam Ingram, Dom Walton, Rob Fender, Chris Done, Misaki Mizumoto, Mariko Kimura, David Tsang, Thomas Dauser, and Javier Garc\'ia for many excellent and helpful conversations. We gratefully acknowledge support from NASA under grants NNX17AF29G and NNX15AU54G. Finally, we would like to thank the Department of Astronomy as the University of Maryland for allowing us to use part of their computational resources on the {\tt yorp} cluster. CSR thanks the UK Science and Technology Facilities Council (STFC) for support.

\textbf{\software{Fenrir}}




\bibliography{ms}

\begin{thebibliography}{}
\expandafter\ifx\csname natexlab\endcsname\relax\def\natexlab#1{#1}\fi

\bibitem[{{Abramowicz} {et~al.}(1980){Abramowicz}, {Calvani}, \&
  {Nobili}}]{Abramowicz+1980}
{Abramowicz}, M.~A., {Calvani}, M., \& {Nobili}, L. 1980, \apj, 242, 772

\bibitem[{{Abramowicz} {et~al.}(1988){Abramowicz}, {Czerny}, {Lasota}, \&
  {Szuszkiewicz}}]{Abramowicz+1988}
{Abramowicz}, M.~A., {Czerny}, B., {Lasota}, J.~P., \& {Szuszkiewicz}, E. 1988,
  \apj, 332, 646

\bibitem[{{Ar{\'e}valo} \& {Uttley}(2006)}]{Arevalo+Uttley2006}
{Ar{\'e}valo}, P., \& {Uttley}, P. 2006, \mnras, 367, 801

\bibitem[{{Bachetti} \& {Huppenkothen}(2018)}]{Bachetti+Hupp2018}
{Bachetti}, M., \& {Huppenkothen}, D. 2018, \apjl, 853, L21

\bibitem[{{Bardeen} {et~al.}(1972){Bardeen}, {Press}, \&
  {Teukolsky}}]{Bardeen+1972}
{Bardeen}, J.~M., {Press}, W.~H., \& {Teukolsky}, S.~A. 1972, \apj, 178, 347

\bibitem[{{Barr} {et~al.}(1985){Barr}, {White}, \& {Page}}]{Barr+1985}
{Barr}, P., {White}, N.~E., \& {Page}, C.~G. 1985, \mnras, 216, 65P

\bibitem[{{Biretta} {et~al.}(2002){Biretta}, {Junor}, \&
  {Livio}}]{Biretta+2002}
{Biretta}, J.~A., {Junor}, W., \& {Livio}, M. 2002, \nar, 46, 239

\bibitem[{{Boyer} \& {Lindquist}(1967)}]{Boyer+Lindquist1967}
{Boyer}, R.~H., \& {Lindquist}, R.~W. 1967, Journal of Mathematical Physics, 8,
  265

\bibitem[{{Caballero-Garcia} {et~al.}(2017){Caballero-Garcia}, {Dovciak},
  {Papadakis}, {Epitropakis}, {Svoboda}, {Kara}, \&
  {Karas}}]{Caballero-Garcia+2017}
{Caballero-Garcia}, M.~D., {Dovciak}, M., {Papadakis}, I., {et~al.} 2017, ArXiv
  e-prints, arXiv:1701.03905

\bibitem[{{Caballero-Garcia} {et~al.}(2018){Caballero-Garcia}, {Papadakis},
  {Dovciak}, {Bursa}, {Epitropakis}, {Karas}, \&
  {Svoboda}}]{Caballero-Garcia+2018}
{Caballero-Garcia}, M.~D., {Papadakis}, I.~E., {Dovciak}, M., {et~al.} 2018,
  ArXiv e-prints, arXiv:1804.03503

\bibitem[{{Cackett} {et~al.}(2014){Cackett}, {Zoghbi}, {Reynolds}, {Fabian},
  {Kara}, {Uttley}, \& {Wilkins}}]{Cackett+2014}
{Cackett}, E.~M., {Zoghbi}, A., {Reynolds}, C., {et~al.} 2014, \mnras, 438,
  2980

\bibitem[{{Chainakun} \& {Young}(2015)}]{Chainakun+Young2015}
{Chainakun}, P., \& {Young}, A.~J. 2015, \mnras, 452, 333

\bibitem[{{Chainakun} \& {Young}(2017)}]{Chainakun+Young2017}
---. 2017, \mnras, 465, 3965

\bibitem[{{Cunningham}(1975)}]{Cunningham1975}
{Cunningham}, C.~T. 1975, \apj, 202, 788

\bibitem[{{Dauser} {et~al.}(2014){Dauser}, {Garc{\'{\i}}a}, {Parker}, {Fabian},
  \& {Wilms}}]{Dauser+2014}
{Dauser}, T., {Garc{\'{\i}}a}, J., {Parker}, M.~L., {Fabian}, A.~C., \&
  {Wilms}, J. 2014, \mnras, 444, L100

\bibitem[{{Dauser} {et~al.}(2012){Dauser}, {Svoboda}, {Schartel}, {Wilms},
  {Dov{\v c}iak}, {Ehle}, {Karas}, {Santos-Lle{\'o}}, \&
  {Marshall}}]{Dauser+2012}
{Dauser}, T., {Svoboda}, J., {Schartel}, N., {et~al.} 2012, \mnras, 422, 1914

\bibitem[{{Emmanoulopoulos} {et~al.}(2014){Emmanoulopoulos}, {Papadakis},
  {Dov{\v c}iak}, \& {McHardy}}]{Emm+2014}
{Emmanoulopoulos}, D., {Papadakis}, I.~E., {Dov{\v c}iak}, M., \& {McHardy},
  I.~M. 2014, \mnras, 439, 3931

\bibitem[{{Fabian} {et~al.}(2015){Fabian}, {Lohfink}, {Kara}, {Parker},
  {Vasudevan}, \& {Reynolds}}]{Fabian+2015}
{Fabian}, A.~C., {Lohfink}, A., {Kara}, E., {et~al.} 2015, \mnras, 451, 4375

\bibitem[{{Fabian} {et~al.}(1989){Fabian}, {Rees}, {Stella}, \&
  {White}}]{Fabian+1989}
{Fabian}, A.~C., {Rees}, M.~J., {Stella}, L., \& {White}, N.~E. 1989, \mnras,
  238, 729

\bibitem[{{Fabian} {et~al.}(2009){Fabian}, {Zoghbi}, {Ross}, {Uttley}, {Gallo},
  {Brandt}, {Blustin}, {Boller}, {Caballero-Garcia}, {Larsson}, {Miller},
  {Miniutti}, {Ponti}, {Reis}, {Reynolds}, {Tanaka}, \& {Young}}]{Fabian+2009}
{Fabian}, A.~C., {Zoghbi}, A., {Ross}, R.~R., {et~al.} 2009, \nat, 459, 540

\bibitem[{{Frederick} {et~al.}(2018){Frederick}, {Kara}, {Reynolds}, {Pinto},
  \& {Fabian}}]{Frederick+2018}
{Frederick}, S.~E., {Kara}, E., {Reynolds}, C.~S., {Pinto}, C., \& {Fabian},
  A.~C. 2018, ArXiv e-prints, arXiv:1802.06056

\bibitem[{{Garc{\'{\i}}a} {et~al.}(2014){Garc{\'{\i}}a}, {Dauser}, {Lohfink},
  {Kallman}, {Steiner}, {McClintock}, {Brenneman}, {Wilms}, {Eikmann},
  {Reynolds}, \& {Tombesi}}]{Garcia+2014}
{Garc{\'{\i}}a}, J., {Dauser}, T., {Lohfink}, A., {et~al.} 2014, \apj, 782, 76

\bibitem[{{Gaskell}(2004)}]{Gaskell2004}
{Gaskell}, C.~M. 2004, \apjl, 612, L21

\bibitem[{{Ghisellini} {et~al.}(2004){Ghisellini}, {Haardt}, \&
  {Matt}}]{Ghisellini+2004}
{Ghisellini}, G., {Haardt}, F., \& {Matt}, G. 2004, \aap, 413, 535

\bibitem[{{Hirotani} \& {Okamoto}(1998)}]{Hirotani+Okamoto1998}
{Hirotani}, K., \& {Okamoto}, I. 1998, \apj, 497, 563

\bibitem[{{Hogg} \& {Reynolds}(2016)}]{Hogg+Reynolds2016}
{Hogg}, J.~D., \& {Reynolds}, C.~S. 2016, \apj, 826, 40

\bibitem[{{Jiang} {et~al.}(2017){Jiang}, {Stone}, \& {Davis}}]{Jiang+2017}
{Jiang}, Y.-F., {Stone}, J., \& {Davis}, S.~W. 2017, ArXiv e-prints,
  arXiv:1709.02845

\bibitem[{{Jiang} {et~al.}(2014){Jiang}, {Stone}, \& {Davis}}]{Jiang+2014}
{Jiang}, Y.-F., {Stone}, J.~M., \& {Davis}, S.~W. 2014, \apj, 796, 106

\bibitem[{{Kara} {et~al.}(2016){Kara}, {Alston}, {Fabian}, {Cackett}, {Uttley},
  {Reynolds}, \& {Zoghbi}}]{Kara+2016}
{Kara}, E., {Alston}, W.~N., {Fabian}, A.~C., {et~al.} 2016, \mnras, 462, 511

\bibitem[{{Kara} {et~al.}(2015){Kara}, {Fabian}, {Lohfink}, {Parker}, {Walton},
  {Boggs}, {Christensen}, {Hailey}, {Harrison}, {Matt}, {Reynolds}, {Stern}, \&
  {Zhang}}]{Kara+2015}
{Kara}, E., {Fabian}, A.~C., {Lohfink}, A.~M., {et~al.} 2015, \mnras, 449, 234

\bibitem[{{Karas} {et~al.}(1992){Karas}, {Vokrouhlicky}, \&
  {Polnarev}}]{Karas+1992}
{Karas}, V., {Vokrouhlicky}, D., \& {Polnarev}, A.~G. 1992, \mnras, 259, 569

\bibitem[{{Kerr}(1963)}]{Kerr1963}
{Kerr}, R.~P. 1963, Physical Review Letters, 11, 237

\bibitem[{{Kotov} {et~al.}(2001){Kotov}, {Churazov}, \&
  {Gilfanov}}]{Kotov+2001}
{Kotov}, O., {Churazov}, E., \& {Gilfanov}, M. 2001, \mnras, 327, 799

\bibitem[{{Martocchia} \& {Matt}(1996)}]{Martocchia+Matt1996}
{Martocchia}, A., \& {Matt}, G. 1996, \mnras, 282, L53

\bibitem[{{Mastroserio} {et~al.}(2018){Mastroserio}, {Ingram}, \& {van der
  Klis}}]{Mastroserio+2018}
{Mastroserio}, G., {Ingram}, A., \& {van der Klis}, M. 2018, \mnras, 475, 4027

\bibitem[{{Matt} \& {Perola}(1992)}]{Matt+Perola1992}
{Matt}, G., \& {Perola}, G.~C. 1992, \mnras, 259, 433

\bibitem[{{McKinney} {et~al.}(2014){McKinney}, {Tchekhovskoy}, {Sadowski}, \&
  {Narayan}}]{McKinney+2014}
{McKinney}, J.~C., {Tchekhovskoy}, A., {Sadowski}, A., \& {Narayan}, R. 2014,
  \mnras, 441, 3177

\bibitem[{{Miniutti} \& {Fabian}(2004)}]{Miniutti+Fabian2004}
{Miniutti}, G., \& {Fabian}, A.~C. 2004, \mnras, 349, 1435

\bibitem[{{Miyamoto} \& {Kitamoto}(1989)}]{Miyamoto+Kitamoto1989}
{Miyamoto}, S., \& {Kitamoto}, S. 1989, \nat, 342, 773

\bibitem[{{Nowak} {et~al.}(1999){Nowak}, {Vaughan}, {Wilms}, {Dove}, \&
  {Begelman}}]{Nowak+1999}
{Nowak}, M.~A., {Vaughan}, B.~A., {Wilms}, J., {Dove}, J.~B., \& {Begelman},
  M.~C. 1999, \apj, 510, 874

\bibitem[{{Page}(1985)}]{Page1985}
{Page}, C.~G. 1985, \ssr, 40, 387

\bibitem[{{Papadakis} {et~al.}(2001){Papadakis}, {Nandra}, \&
  {Kazanas}}]{Papadakis+2001}
{Papadakis}, I.~E., {Nandra}, K., \& {Kazanas}, D. 2001, \apjl, 554, L133

\bibitem[{{Pariev} \& {Bromley}(1998)}]{Pariev+Bromley1998}
{Pariev}, V.~I., \& {Bromley}, B.~C. 1998, \apj, 508, 590

\bibitem[{{Reis} \& {Miller}(2013)}]{Reis+Miller2013}
{Reis}, R.~C., \& {Miller}, J.~M. 2013, \apjl, 769, L7

\bibitem[{{Reynolds}(2014)}]{Reynolds2014}
{Reynolds}, C.~S. 2014, \ssr, 183, 277

\bibitem[{{Reynolds} \& {Begelman}(1997)}]{Reynolds+Begelman1997}
{Reynolds}, C.~S., \& {Begelman}, M.~C. 1997, \apj, 488, 109

\bibitem[{{Reynolds} \& {Fabian}(2008)}]{Reynolds+Fabian2008}
{Reynolds}, C.~S., \& {Fabian}, A.~C. 2008, \apj, 675, 1048

\bibitem[{{Reynolds} {et~al.}(1999){Reynolds}, {Young}, {Begelman}, \&
  {Fabian}}]{Reynolds+1999}
{Reynolds}, C.~S., {Young}, A.~J., {Begelman}, M.~C., \& {Fabian}, A.~C. 1999,
  \apj, 514, 164

\bibitem[{{Ross} \& {Fabian}(1993)}]{Ross+Fabian1993}
{Ross}, R.~R., \& {Fabian}, A.~C. 1993, \mnras, 261, 74

\bibitem[{{Shakura} \& {Sunyaev}(1973)}]{Shakura+Sunyaev1973}
{Shakura}, N.~I., \& {Sunyaev}, R.~A. 1973, \aap, 24, 337

\bibitem[{{Shapiro}(1964)}]{Shapiro1964}
{Shapiro}, I.~I. 1964, Physical Review Letters, 13, 789

\bibitem[{{Stella}(1990)}]{Stella1990}
{Stella}, L. 1990, \nat, 344, 747

\bibitem[{{Taylor} \& {Reynolds}(2018)}]{Taylor+Reynolds2018}
{Taylor}, C., \& {Reynolds}, C.~S. 2018, \apj, 855, 120

\bibitem[{{Uttley} {et~al.}(2014){Uttley}, {Cackett}, {Fabian}, {Kara}, \&
  {Wilkins}}]{Uttley+2014}
{Uttley}, P., {Cackett}, E.~M., {Fabian}, A.~C., {Kara}, E., \& {Wilkins},
  D.~R. 2014, \aapr, 22, 72

\bibitem[{{Uttley} \& {McHardy}(2001)}]{Uttley+McHardy2001}
{Uttley}, P., \& {McHardy}, I.~M. 2001, \mnras, 323, L26

\bibitem[{{Uttley} {et~al.}(2005){Uttley}, {McHardy}, \&
  {Vaughan}}]{Uttley+2005}
{Uttley}, P., {McHardy}, I.~M., \& {Vaughan}, S. 2005, \mnras, 359, 345

\bibitem[{{Wilkins} {et~al.}(2016){Wilkins}, {Cackett}, {Fabian}, \&
  {Reynolds}}]{Wilkins+2016}
{Wilkins}, D.~R., {Cackett}, E.~M., {Fabian}, A.~C., \& {Reynolds}, C.~S. 2016,
  \mnras, 458, 200

\bibitem[{{Wilkins} \& {Fabian}(2011)}]{Wilkins+Fabian2011}
{Wilkins}, D.~R., \& {Fabian}, A.~C. 2011, \mnras, 414, 1269

\bibitem[{{Wilkins} \& {Fabian}(2012)}]{Wilkins+Fabian2012}
---. 2012, \mnras, 424, 1284

\bibitem[{{Wilkins} {et~al.}(2017){Wilkins}, {Gallo}, {Silva}, {Costantini},
  {Brandt}, \& {Kriss}}]{Wilkins+2017}
{Wilkins}, D.~R., {Gallo}, L.~C., {Silva}, C.~V., {et~al.} 2017, \mnras, 471,
  4436

\bibitem[{{Wu} \& {Wang}(2007)}]{Wu+Wang2007}
{Wu}, S.-M., \& {Wang}, T.-G. 2007, \mnras, 378, 841

\bibitem[{{Zoghbi} {et~al.}(2012){Zoghbi}, {Fabian}, {Reynolds}, \&
  {Cackett}}]{Zoghbi+2012}
{Zoghbi}, A., {Fabian}, A.~C., {Reynolds}, C.~S., \& {Cackett}, E.~M. 2012,
  \mnras, 422, 129

\bibitem[{{Zoghbi} {et~al.}(2014){Zoghbi}, {Cackett}, {Reynolds}, {Kara},
  {Harrison}, {Fabian}, {Lohfink}, {Matt}, {Balokovic}, {Boggs}, {Christensen},
  {Craig}, {Hailey}, {Stern}, \& {Zhang}}]{Zoghbi+2014}
{Zoghbi}, A., {Cackett}, E.~M., {Reynolds}, C., {et~al.} 2014, \apj, 789, 56

\end{thebibliography}
\end{document}